\RequirePackage{fix-cm}
\documentclass[smallextended]{svjour3}       
\smartqed  
\usepackage{graphicx}
\usepackage{subcaption}
\usepackage{textcomp}
\usepackage{biblatex}
\usepackage{multirow}
\usepackage{arydshln}
\usepackage{longtable}
\usepackage{float}

\graphicspath{ {./images/} }
\usepackage[table,dvipsnames]{xcolor}
\usepackage{pgfplots}
\usepackage{tcolorbox}
\tcbuselibrary{listings,theorems,skins}
\newcolumntype{Y}{>{\centering\arraybackslash}X}
\usepackage{breakcites}
\usepackage{footnote}
\begin{document}

\title{A Pragmatic Approach for Hyper-Parameter Tuning in Search-based Test Case Generation
}

\author{Shayan Zamani         \and
        Hadi Hemmati 
}


\institute{Shayan Zamani \at
              Department of Electrical and Computer Engineering, Schulich School of Engineering \\
              University of Calgary \\
              \email{shayan.zamani1@ucalgary.ca}           
           \and
           Hadi Hemmati \at
              Department of Electrical and Computer Engineering, Schulich School of Engineering \\
              University of Calgary\\
              \email{hadi.hemmati@ucalgary.ca} 
}

\date{Received: date / Accepted: date}

\maketitle

\begin{abstract}
Search-based test case generation, which is the application of meta-heuristic search for generating test cases, has been studied a lot in the literature, lately. Since, in theory, the performance of meta-heuristic search methods is highly dependent on their hyper-parameters, there is a need to study hyper-parameter tuning in this domain. In this paper, we propose a new metric (``Tuning Gain''), which estimates how cost-effective tuning a particular class is. We then predict ``Tuning Gain'' using static features of source code classes. Finally, we prioritize classes for tuning, based on the estimated ``Tuning Gains'' and spend the tuning budget only on the highly-ranked classes. 
To evaluate our approach, we exhaustively analyze 1,200 hyper-parameter configurations of a well-known search-based test generation tool (EvoSuite) for 250 classes of 19 projects from benchmarks such as SF110 and SBST2018 tool competition. We used a tuning approach called Meta-GA and compared the tuning results with and without the proposed class prioritization. The results show that for a low tuning budget, prioritizing classes outperforms the alternatives in terms of extra covered branches (10 times more than a traditional global tuning). 
In addition, we report the impact of different features of our approach such as search space size, tuning budgets, tuning algorithms, and the number of classes to tune, on the final results.



\end{abstract}

\section{Introduction}
\label{intro}

Software testing is one of the most common approaches to verify a software system, that is being used in practice. Testing is the process of executing a source code with different inputs and scenarios and in different contexts to make sure its actual behavior matches its expected behavior, in all those contexts and scenarios.  There has been many solutions proposed and developed in recent years to automate the process of test case generation, among which search-based techniques have been quite effective in many systems and domains~\cite{sbse,sbst,sbst_survey}. Currently, there are several publicly available search-based test generation (SBTG) tools such as EvoSuite for unit testing Java programs that are regularly being maintained and improved~\cite{evosuite}.

In general, SBTG techniques reformulate test generation problems, e.g. maximizing branch coverage of unit tests, to an objective function and employ a meta-heuristic search technique to optimize the objective. 
Evolutionary Algorithms, like Genetic Algorithm (GA), are among the most common techniques that have been used in SBTG.
The performance of a GA-based SBTG approach (e.g. in terms of code coverage) can be highly impacted by the choice of the GA's chromosome encoding, its objective function, and the GA's hyper-parameters (such as population size, crossover rate, mutation rate, etc.)~\cite{GAbyGA,andrea,bookchapter_testing}.

Therefore, tuning the hyper-parameters to find an optimal configuration could potentially improve the effectiveness of SBTG, significantly. In general, there are many parameters to be tuned for a GA. For example, Grefenstette used six parameters (Population Size, Crossover Rate, Mutation Rate, Generation Gap, Scaling Window, and Selection Function) to tune the GA~\cite{controlGA}. However, in another paper, 19 different operators or parameters are listed that contribute to the performance of a given GA~\cite{GAbyGA}.

Although the effectiveness of tuning is being studied in other areas frequently~\cite{review1,review2,review3,Feldt:2000}, there are not many successful reports of tuning techniques in SBTG. For example, Arcuri et al. tried to find a tuned setting for EvoSuite, but the resulting branch coverage after tuning was less than the default configuration's results~\cite{andrea}. 

Regardless of the effectiveness of tuning, in general, tuning is always an overhead. Therefore, any tuning study should carefully investigate its cost, as well. In the context of SBTG tuning, this is particularly important, since (a) evaluating each combination of hyper-parameter values per class in SBTG means running an SBTG tool like EvoSuite on that class, which is relatively expensive and (b) a typical tuning of an SBTG tool would require all classes of a project to be considered for tuning, each with many combinations of hyper-parameter values.
Therefore, a typical tuning can become quite ineffective (with a limited budget) or impractical, for real-world large scale projects with many classes.

The above challenges motivate us to design a pragmatic tuning method for SBTG.  The key goal is to design a cost-effective tuning method that spends the limited tuning budget wisely across the project, assuming that only a small subset of classes are worth tuning. In other words, if we only focus on a small subset of classes, we can afford to allocate more tuning budget, for each class within the selected subset.

One challenge here is to find the right subset of classes that worth tuning. Previous studies~\cite{cov_predict,adaptive,branchexpect} have shown that software testing complexity can be estimated using source code metrics. Therefore, we propose to use a set of static source code metrics to predict the ``worthiness'' of classes within a project to allocate the limited tuning budget only to the highly ranked classes and not the entire project.

Therefore, the key idea of our approach is to first estimate the cost- effectiveness of tuning per individual class, using static source code metrics. Then, these classes are prioritized based on the predication. Then the limited tuning budget is only assigned to the highest ranked classes and the default configuration is used for the rest.

To assess our proposal, we study five research questions, as follows:



In our research question 1, we define a measure for the cost-effectiveness of tuning in SBTG (``Tuning Gain'') as the product of maximum effectiveness (how much extra coverage one can gain by an optimal tuning compared to average) and the difficulty of tuning (how complex it is to find the optimal tuning), per class. We then estimate the ``Tuning Gain'' metric by a regression model on some static metrics of the source code, per class (e.g. line of codes, McCabe's Complexity, Halstead Complexity, etc.). We show that the regression model can predict a close estimate of class rankings with respect to their ``Tuning Gain''.

In research question 2, initially, we evaluate the effectiveness of our class prioritization-based tuning approach both in terms of the proposed class ordering method and the search algorithm used during tuning, independently. We then compare the cost-effectiveness of our approach with a global tuning (i.e, one tuning for all classes), random class-level tuning, basic (evenly distributed) class-level tuning, and no tuning (default). We show that given the same budget, our approach is superior over all alternatives. 

In research question 3, we study the effect of the number of classes selected for tuning, in our approach. We discuss that if we consider the entire project then most likely no more than 20\% of classes are needed for tuning. But if one uses a subset of only highly potential classes for tuning, per project, a higher cut-off is needed (around 60\% in our case).  

The next two research questions try two different ways to increase the tuning budget per class, as follows: 

In research question 4, we reduce the size of configurations search space (number of hyper-parameters to tune), which results in exploring more instances in the search space, per class. We show that a smaller search-space can result in coverage as good as the full search-space, but not better, given the same budget. 

Finally, in research question 5, we study the effect of search algorithms within our approach. To understand their effect, we increase the overall tuning budget to let the search algorithms show their potentials. The observation is that with a small budget, when not all classes can be allocated enough tuning budget, prioritization plays a significant role and the prioritized meta-GA approach is the dominant approach. However, as the budget increases the rates of increments in coverage using our approach and others are not much different.

The main contributions of this paper are as follows:
\begin{itemize}

\item Introducing a new metric (``Tuning Gain'') to measure the potential cost-effectiveness of a tuning method, per class;

\item Accurately (with an AUC of 92\%) estimating the ``Tuning Gain'' using static source code metrics, per class;

\item Proposing a class-prioritization-based tuning method that allocates the tuning budget only to a subset of classes with the highest predicted ``Tuning Gain'', which improved the global tuning results by 10 times, for low-tuning budgets;

\item Running a large-scale study with over 3 million configuration-class pairs, to evaluate and compare the cost-effectiveness of the proposed approach with traditional tuning methods.

\item Empirically investigating the effect of ``search method'', ``tuning subset'' and ``search space'' sizes, within our proposed approach.

\end{itemize}

The rest of this paper is organized as follows: 
In section \ref{motive}, we explain the motivation of this study. Section \ref{background} explains the required background and related work. Our class prioritization technique is elaborated in section \ref{method}. Then, in section \ref{empirical}, the experiment setup, design, and results are explained in detail. Section \ref{future} briefly explains the future work. Finally, section \ref{conclusion} concludes the study. 

\section{Motivation}
\label{motive}
Unlike many similar domains, such as applications of machine learning, where hyper-parameter tuning is a norm~\cite{Feurer2019}, tuning is not largely used in SBTG studies and practice. The main reasons are 1) SBTG tuning is expensive and 2) the benefits are not significant compared to a default (or even a random~\cite{khodam}) setup. 

Arcuri and Fraser~\cite{andrea} demonstrated that although for some individual classes there is a significant improvement in coverage after tuning, the default configuration covers more total branches than the tuned configuration, across the entire project, when one tunes all classes of a project together.

Moreover, Zamani and Hemmati~\cite{khodam} showed for a collection of classes even random configurations often cover few more or at least as many branches as a default configuration does. They illustrated that if we tune classes individually rather than all together (the entire project as a whole), the improvements in coverage would be more significant. They also reported that, in their study, around 80\% of classes do not need tuning, and they covered the same branches regardless of the tool configuration.

The above challenges motivate us to design a pragmatic tuning method for SBTG.  The key goal is to spend the limited tuning budget wisely across the project, assuming that only a small subset of classes are worth tuning. In other words, if we only focus on a small subset of classes, we can afford to allocate more tuning budget, for each class within the selected subset. Therefore, the challenge here is to find the right subset of classes that worth tuning.

Previous studies~\cite{cov_predict,adaptive,branchexpect} have shown that software testing complexity can be estimated using source code metrics. Therefore, we propose to use a set of static source code metrics to predict the ``worthiness'' of classes within a project to allocate the limited tuning budget only to the highly ranked classes and not the entire project. 


\section{Background and Related Work}
\label{background}
In this section, we briefly explain the background required to understand the rest of the paper, in terms of metaheuristic tuning methods and source code metrics.

\subsection{Search-based Test Case Generation}

Search-based test case generation (SBTG) is an automated test generation strategy that reformulates the test generation problem to an optimization problem, where the goal is to generate test data that maximizes a test adequacy criterion (such as statement and branch coverage) ~\cite{sbst_survey}. There are many metaheuristic search algorithms that can be used for this optimization. 

\begin{figure}[htbp]
\centering
\includegraphics[width=\textwidth]{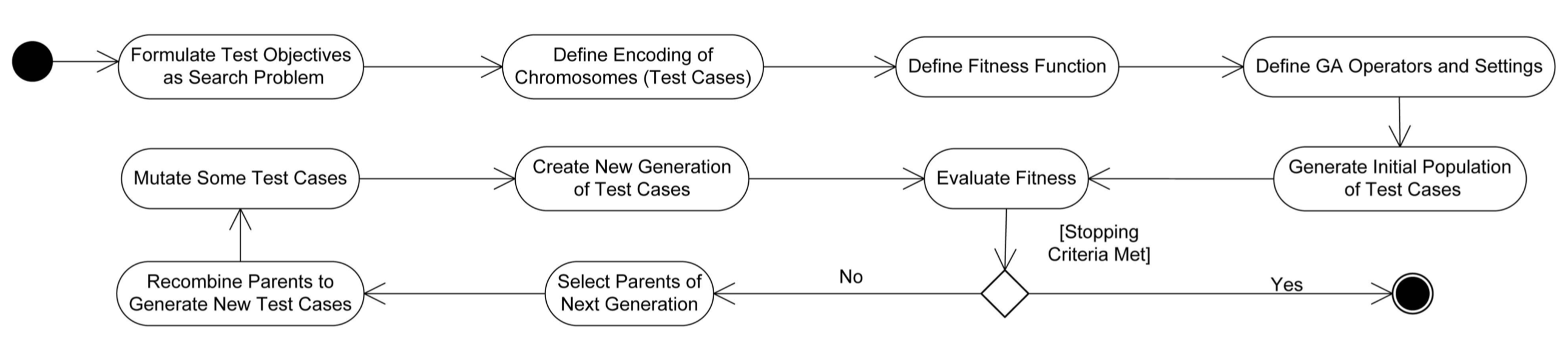}
\caption{Search-based test case generation (SBTG) using Genetic Algorithms (GA)~\cite{sbst_survey}.}
\label{fig:sbst}
\end{figure}

Figure \ref{fig:sbst} illustrates the most common SBTG strategy, which employs genetic algorithms. 

In Genetic Algorithms, a population of random solutions is selected and evolved with crossover and mutation operations. Next populations will contain the solutions with the highest fitness and solutions generated from the recombination of the previous population~\cite{sbst,harmancurrent}. 
    
There are different test case generation tools using search-based methods~\cite{JTExpert,DOTgEAr}. EvoSuite~\cite{evosuite}, for example, use a genetic algorithm to evolve the entire test suite as a whole rather than evolving test cases, individually.

\subsection{Hyper-parameter Tuning Methods}

Hyper-parameter tuning is the practice of finding values for input parameters of an algorithm, so that its performance becomes optimal. In parameter tuning, once the parameter values are selected, they remain fixed during the run. The intuitive approach to parameter tuning is to experiment with all different combinations of parameters' values and select the one with the best performance. However, this exhaustive tuning is very time consuming, and sometimes impossible (e.g., if the inputs types are continuous). Therefore several tuning approaches have been proposed to optimize this process ~\cite{Eiben2003}. The main tuning methods employed in the software engineering literature are the followings~\cite{improvedSecBug}:

\begin{enumerate}
    \item \underline{Grid Search}~\cite{algorithms_for_HPO,classification_tuning} where first a handful of values for each hyper-parameter is selected and then different combinations of those hyper-parameter values are exhaustively evaluated. The main issue with this brute-force approach is ``curse of dimensionality'', and we may lose much information by limiting the search space to discrete values.
    
    \item \underline{Random Search}~\cite{random_bengio}, where the solutions from a search space, are selected, stochastically out of a probability distribution. This method does not use the information from previous trials to guide the search, but it is usually more effective than a grid search when the same budget is given to them.
    
    \item \underline{Meta-heuristic Search}, such as Genetic Algorithm ~\cite{goldberg2006genetic} are a variant of random search, where a population of solutions is evaluated and the next generations are formed based on the elite solutions of the last generation. Other used meta-heuristic search algorithms for tuning are: Simulated Annealing, which is basically an evolutionary algorithm in which the population size is one, and Hill Climbing, which is a variation of simulated annealing, where there is no option to escape from the local optima~\cite{simulated_annealing}.
    
    \item \underline{Bayesian Optimization}~\cite{boa,thomas} which is a way to model the objective function, statistically. After a few random observations in the search space, it uses an acquisition function to guide its search to the most informative observations. The information gain of points in the search space is calculated based on the posterior probability distribution of the already observed points. This method only fits to problems where the hyper-parameters are continuous.
    
    \item \underline{Differential Evolution (DE)}~\cite{differential_evolution}. In evolutionary algorithms, attributes of a solution are mutated randomly. However, in DE, before mutation, we first pick three random solutions from a list called frontier. Then, attributes of the solution are mutated based on the extrapolation of the three picked solutions. This method recently got researchers' attention in software engineering~\cite{SMOTUNED,topicModelling,DE+RF,necessary,Fast-and-Frugal}.
    
\end{enumerate}  

In the evolutionary algorithm literature, there are also some other methods that are specialized for tuning meta-heuristics:
    
\begin{enumerate}
  \setcounter{enumi}{5}
  \item F-Race~\cite{frace}: The main concept in this tuning method is a race in which the poor configurations in terms of fitness are dropped from the search space and they will no longer be evaluated. Therefore, the focus of the search will only be on promising configurations. Racing is repeated until the best configuration is found. This method uses the statistical Friedman two-way analysis of variance by ranks to eliminate candidates with a low fitness score.
  
  \item Response Surface Methodology~\cite{rsm} evaluates some critical spots in the search space based on the range of variation that parameters exhibit. Then, based on the fitness scores and the gradients of fitness in each direction of the search space, it moves toward the answer with higher fitness. This method was used in the SBST context for tuning and was said to be ineffective to find an optimal value for EvoSuite tool~\cite{andrea}.
\end{enumerate}

\subsection{Hyper-parameter Tuning in Software Engineering}
\label{related_work}

The effectiveness of tuning in search-based software engineering has been studied in several papers. For example, in the Software Product Lines optimization problem, in which the objective is to satisfy stakeholder needs, a comparison of the results of two search methods showed that tuning hyper-parameters can have an impact on the results~\cite{replicate}.

However, there have been very few studies on tuning SBTG hyperparameters, so far. Arcuri et al.~\cite{andrea} conducted three case studies on 20 handpicked Java classes. They first observed that tuning has a high impact on the coverage values for individual classes. Then, they extended their study to 609 classes from 10 random projects, and after tuning the whole set of classes, they came up with a tuned configuration by RSM method that ended up with lower coverage for the whole classes compared to the default configuration. Therefore, they concluded that although tuning is promising in single classes, it is not scalable to the projects~\cite{andrea}. Later, the replication of this paper by Kotelyanskii et al. used Sequential Parameter Optimization Toolbox (SPOT) to tune SBST tool and the results confirmed that project-level tuning can not improve the coverage that can be derived by the default configuration for EvoSuite~\cite{replication}. 

Parameter tuning has also been shown to be effective in problems that employ machine learning algorithms. 87\% of classification methods used in software engineering necessitate parameter tuning for at least one parameter~\cite{classification_tuning,necessary}. For example, tuning machine learning was shown to be effective in software effort estimation~\cite{MLtuning,effortEstim}, software development defect predication~\cite{defect,DE+RF} and clone detection tools~\cite{cloneDetection}. Furthermore, in topic modelling, Panichella et al. showed that the performance of Latent Dirichlet Allocation (LDA) in SE tasks would be improved with the right selection of four hyper-parameters corresponding to LDA~\cite{topicModel_GA}.

Additionally, there are many studies in the field of software engineering where different hyper-parameter optimizers are compared with each other. Agrawal et al. \cite{dodge}, for example, compared defect prediction results when different hyper-parameter optimizers including differential evolution-based algorithms~\cite{SMOTUNED,DE+RF} where used. In the release planning problem, Zhang et al. found that using hyper-heuristic methods can lead the search to more scalable solutions~\cite{ruhe_release_planning}. LDA topic modelling performance was also improved when using differential evolution~\cite{topicModelling} in comparison to the genetic algorithm that was used previously~\cite{topicModel_GA}.

In this empirical study, we use methods from Grid Search (in RQ1) and Meta-heuristics Search methods (in RQ2 and RQ3). More precisely, we use an evolutionary technique called Meta-GA that uses a GA on top of another GA. This method is initially defined by Grefenstette~\cite{controlGA} and has evolved since then~\cite{GAbyGA} and is shown to be effective in tuning the meta-heuristics' parameters~\cite{comparison}. Meta-GA's chromosomes are GAs with different configurations, and its fitness function is the performance of each GA (here, it is the coverage of each GA). In other words, Meta-GA searches a space of GA configurations in order to identify an efficient GA. We compare the results of Meta-GA with Random Search, and Differential Evolution strategies, as well.

\subsection{Static Source Code Metrics}

Given our motivation, we need to estimate which classes are likely to gain more coverage after tuning. We look at classes' static features that can be extracted without execution. We used 4 different tools to extract these static metrics for Java code, as explained in this section. Considering that our study leverages a Java-based SBTG tool (Evosuite), we focus on metrics that are introduced for Java, as follows:

\begin{enumerate}
    \item \underline{Package-level metrics:}
    These features measure how well the Java packages are designed by traversing the Java file directories. JDepend~\cite{jdepend} generates 8 package-level static measures including Number of Classes, Number of Abstract Classes, Afferent Couplings, Efferent Couplings, Abstractness, Instability, Distance from the Main Sequence and Package Dependency Cycles. Since most of these metrics are quite self-descriptive due to the lack of space, we do not define them here. To get a full definition, one can read the tool description guide~\cite{jdepend}.

    \item \underline{Object-oriented metrics:}
    Object-oriented metrics~\cite{cov_predict} represent the class-level properties and are calculated using the CK Tool, which is developed for Java language and is freely available on GitHub~\cite{cktool}.
    There are 39 static metrics that CK Tool measures for a given Java class and their description is provided on the tool's website. We used all 39 metrics such as: RFC (Response for a Class), which is the Number of unique method invocations in a given Java class, WMC (Weight Method Class) or McCabe's complexity~\cite{mccabe,adaptive}, which is the number of branch instructions in a class, LOC (Lines of code), which is the number of lines code, etc. 
    
    \item \underline{Java keywords:}
    Ignoring the occurrences in the comments, each measure in this category counts the frequency of a Java keyword (total of 52 keywords) per class under study~\cite{silly}.

    \item \underline{Halstead complexity metrics:}
    The Halstead Complexity Metrics are calculated based on the number of operators and operands within a class~\cite{halstead}. They include Program Vocabulary, Volume, Difficulty and Effort. These metrics were used previously in the literature~\cite{cov_predict} for measuring the complexity of Java classes.

\end{enumerate}

The summary of the metrics per category is provided in Table \ref{table:1}.

\begin{table}[h!]
\centering
\caption{Number of static metrics, per category.}
\begin{tabular}{||lr||} 
\hline
 Static Metric Category & Number of Metrics \\ [0.5ex] 
 \hline\hline
 Package-level & 8\\
 Class-level & 39\\
 Keyword Counts & 52\\
 Halstead & 6\\
\hline \hline
Total & 105\\
\hline
\end{tabular}
\label{table:1}
\end{table}

In the context of using static metrics to predict testing properties of classes, Grano et al. recently published a paper that predicts the coverage values of Java classes using machine learning algorithms like Random Forest Regressor~\cite{cov_predict}. They picked 79 different static values for a set of more than 3,000 Java classes and predicted the coverage for them using different algorithms and different search budgets for two SBST tools, namely, EvoSuite and Randoop. The results were promising and they could predict the values for coverage with 0.15 mean absolute error for EvoSuite. We employed their paper to come up with a good set of static metrics that are used in the literature. In this paper, however, we used more static features and leveraged them to predict the ``Tuning Gain'' and not the coverage values.

Regarding the complexity metrics, previous papers introduced and defined static metrics that are highly correlated with the complexity of testing. These metrics are based on the Markov model of the program, and they show how difficult it is for an automated test generating tool to cover branches within a class. For instance, in one study ~\cite{branchexpect} Branch Coverage Expectation employs a fixed matrix of probabilities for its Markov model, whereas in another study ~\cite{adaptive} the expected number of visits of a branch is modifying the matrix each time. In our study, we included all of the metrics used in these papers.

\section{Class-level Prioritization for SBTG Tuning}
\label{method}

Our proposed tuning method is a class-level tuning, in which we do not find one single configuration for the entire set of classes. Instead, we optimize the SBTG's hyper-parameters per class. However, according to the related work~\cite{khodam}, tuning every single class is a waste of budget because around 80\% of classes in a project will not be benefited by tuning. Therefore, the goal of our tuning approach is to allocate the limited tuning budget only to a subset of classes that are predicted to get the most benefits from tuning (i.e. having the highest predicted ``Tuning Gain''). In other words, for  classes that are not selected for tuning, we use the default configuration, but SBTG may have a different configuration setup for the tuned classes.   

\begin{figure*}[h]
\includegraphics[width=\textwidth]{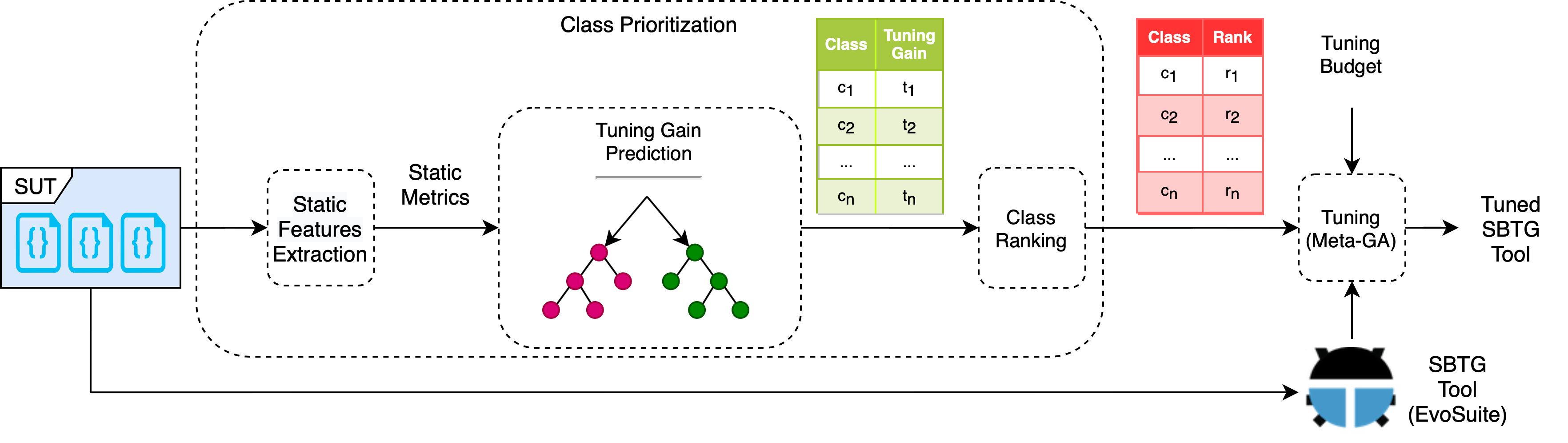}
\caption{The Class Prioritization-based Tuning Approach for SBTG.}
\centering
\label{Fig:proc}
\end{figure*}

As discussed, the key element of our tuning method is to measure the ``worthiness'' of each class, with respect to tuning. ``Tuning Gain'' is the metric we design in this study for this purpose.

In general, in the context of SBTG tuning, classes exhibit three different behaviour types:
\begin{itemize}
    \item Most classes have the same coverage for all configurations. The ``Tuning Gain'' for these classes must be zero. 
    \item Some classes show higher coverage for just a very few configurations. In other words, it is difficult for a tuning method to find those configurations and show its benefit for that class. In other words, spending the tuning budget on these classes is only justified if the tuned configuration achieves quite a high coverage (the increased effectiveness worth the extra cost). So ``Tuning Gain'' for these classes should be typically low but non-zero, unless for a few classes that the coverage gain is very high.
    \item The most interesting classes, for the tuning purpose, are the ones with high variance of coverage and many possibilities (easily detectable) of (near) optimum solutions. ``Tuning Gain'' for these classes should be high. 
\end{itemize}{}

Given the above, we formally define ``Tuning Gain'' as:
\begin{equation}
Tuning Gain =  Variation_{Coverage}  \times Sparsity_{Best Config}
\end{equation}
in that ``Variation of Coverage'' is the standard deviation of the number of covered branches per class across different configurations and repetitions (when using different seeds in EvoSuite per configuration), and the ``Best Configuration Sparsity'' is the difference between the maximum coverage and the median coverage, per class (again across different configurations and repetitions). 

It can be inferred from the high Variation that spending tuning budget on the class is worthy because most likely the coverage using the tuned configuration is different than the coverage using the default setup. The high Sparsity on the other hand, means that the best configuration results in much higher coverage than the coverage using the default setup (assuming that the default is close to median). Therefore, the higher ``Tuning Gain'', the more likely that tuning will find an optimum (or a near-optimum) configuration that results in much better coverage than the default setup's resulting coverage. 

The defined ``Tuning Gain'' can be seen as a measure of cost-effectiveness for SBTG tuning since it not only takes into account the potential effectiveness of a class tuning (using the Sparsity metric) but also indirectly considers the potential cost of tuning using the Variation metric. Note that a low Variation means most solutions are quite similar and thus finding the optimum most likely will take a long time.

Having the ``Tuning Gain'' metric now we can propose our class prioritization approach for tuning, as follows:

Our solution is to estimate the tuning gain using static metrics. To predict it, we build a regression model in which the dependent variable is ``Tuning Gain'' for each class, and the independent variables are the 105 static metrics of the same class.

Therefore, each time that one wants to tune a set of classes, they need to first estimate the tuning gain of classes using our model, then rank the classes based on their tuning gain in descending order. After that, they can narrow down the tuning on as many classes as they want from the top of the list, based on the budget that they have allocated to tuning. Figure \ref{Fig:proc} illustrates this process in details.

\section{Empirical Study}
\label{empirical}
In this section, we explain the design and results of our experiments for evaluating the proposed SBTG tuning method.
\subsection{Research Questions}
\begin{itemize}

\item {\bf RQ1:} {\it How accurately static source code metrics can predict the cost-effectiveness of SBTG tuning, per class?}\\
The idea of this RQ is to evaluate the prediction of ``Tuning Gain'' as a measure of SBTG tuning cost-effectiveness, per class, using a regression model. We also look at the source code metrics' effect and see if we can reduce our feature set without loss of prediction power.

\begin{itemize}
\item {\bf RQ1.1:} {\it How accurately static source code metrics of a class can predict its ``Tuning Gain'', using a random forest regression?}\\
In this sub-RQ, we apply a Random Forest Regression model on static source code metrics of classes to predict their ``Tuning Gain'' and compare the results with several baselines, including the optimal solution.

\item {\bf RQ1.2:} {\it Which static source code metrics are more powerful for ``Tuning Gain'' prediction?}\\
Here we go deeper into the RQ1.1 results and see if we really need all 105 collected static source code metrics for ``Tuning Gain'' prediction and if not, which subset is effective enough.  

\end{itemize}

\item {\bf RQ2:} {\it How cost-effective is our class prioritization technique for SBTG tuning, compared to traditional tuning methods?}\\
Our tuning method consists of two components: 1) a class prioritization method and 2) a search algorithm. Thus, in this RQ, we first evaluate the effectiveness of both components, independently, in terms of code coverage of the generated test cases. Then the cost of our approach (in terms of execution time) will be examined and the overall cost-effectiveness of the approach will be discussed.

\begin{itemize}
\item {\bf RQ2.1:} {\it How effective is our class prioritization method compared to a random ordering of classes, in terms of their achieved extra code coverage?}\\
In this sub-RQ, we focus only on the prioritization component of our approach and compare the final code coverage results, when sorted and unsorted classes are tuned using meta-GA and random search.

\item {\bf RQ2.2:} {\it How effective is the MetaGA search compared to alternative search methods, within the hyper-parameters' search-space, in terms of their achieved extra code coverage?}\\
In this sub-RQ, we keep the class prioritization approach constant and compare the effect of different search techniques i.e. random search, Hill climbing, Meta-GA, and Differential Evolution, in terms of extra coverage they can provide in comparison to default.

\item {\bf RQ2.3:} {\it What is the cost of our SBTG tuning, in practice?}\\
In this sub-RQ, we focus on the overhead cost of tuning and discuss its benefits over running EvoSuite without tuning for more time.

\end{itemize}

\item {\bf RQ3:} {\it What is the effect of the number of classes selected for tuning on the overall effectiveness of our approach?}\\
In this RQ, we want to observe given the same budget for tuning, whether tuning more classes (exploration), or searching a few classes in more depth (exploitation) is more effective, in terms of achieving extra code coverage.

\item {\bf RQ4:} {\it Does reducing the search space, by selecting a subset of hyper-parameters, improve the effectiveness of tuning?}\\
In this RQ, we evaluate the effect of search space size on the tuning effectiveness (achieving higher coverage) and speed (reaching to the high coverage faster).

\begin{itemize}
    \item {\bf RQ4.1:} {\it Which hyper-parameters are safer to discard, if we reduce the search space to a subset of hyper-parameters?}\\
    In this sub-RQ, we assess the importance of tuning each hyper-parameter, individually, using a feature selection analysis on the training set.
    
    \item {\bf RQ4.2:} {\it How does eliminating less important hyper-parameters affect the effectiveness of tuning?}\\
    In this sub-RQ, we investigate whether the same tuning method with the same budget can reach to a given coverage level faster, if we reduce its search space or not.
    
\end{itemize}

\item {\bf RQ5:} {\it What is the effect of tuning budget on the effectiveness of our approach?}\\
In this RQ, we provide much higher tuning budget to our approach and an alternative (a class prioritization method followed by random search for tuning) and study their relative coverage changes as the budget increases, to understand what technique may be useful for lower budget and which one for higher budget.

\end{itemize}

\subsection{Subject Under Study}
The subjects under our study consist of three categories of Java classes. The first category contains all classes from three Java projects. These three projects were randomly picked from the SF110 benchmark, which has been used several times in the past in the SBTG tuning context. We were careful to select a representative set within the benchmark in terms of the number of classes~\cite{benchmark}. Given that small projects with a low number of classes can be trivial for the SBST tool, even without tuning, we selected projects with more classes than the median number of classes in the benchmark.

The second category, from which the Java classes were picked, contains 10 random projects from the SF110 benchmark (with no overlap with projects in the first category). But this time we only select up to five classes per project. The goal is increasing diversity of included classes by selecting them from different projects, but at the same time, keeping the total experiment's size manageable (note that we need an exhaustive search within the hyper-parameters space, per class. Thus, including all classes from all projects was not practical).

The subset selection was random, but we filtered out the classes that do not use the whole search budget. We wanted to focus on the classes that are not easy for the SBTG tool to cover. Some projects did not have 5 classes that use the whole search budget, so their subset includes fewer classes.

Some of our subjects under study from SF110 have also been used in another SBTG tuning study \cite{panichella2017automated}; 27 out of 346 Java classes that they studied from SF110, are common with ours.

The third category of the Java classes uses the same strategy as the second one (to select a set of five classes), but this time from six projects that were under study in the previous SBTG tool competitions~\cite{toolcompetition}. We wanted to make sure that our results were not biased to the SF110 projects.

In total, we conducted our experiments on 250 classes, from 19 different projects, from three different categories. The name of the projects and the number of included classes per project are available in Table \ref{table:2}.

\begin{table}[h!]
\centering
\caption{Description of projects under study.}
\resizebox{8cm}{!}{\begin{tabular}{||clc||} 
\hline
 Category & Project & Number of Classes \\ [0.5ex] 
 \hline\hline
 \multirow{3}{*}{SF110~\cite{benchmark}}
 & JSecurity & 85\\
 & Geo-Google & 56\\
 & JOpenChart & 36\\
\hline 
\multirow{10}{*}{SF110}
 & gfarcegestionfa & 4 \\
 & mygrid & 5\\
 & petsoar & 4\\
 & lhamacaw & 5\\
 & db-everywhere & 4\\
 & a4j & 4\\
 & freemind & 5\\
 & heal & 5\\
 & saxpath & 3\\
 & biblestudy & 4\\
\hline
\multirow{6}{*}{Tool Competition~\cite{toolcompetition}}
 & webmagic & 5\\
 & dubbo & 5\\
 & jsoup & 5\\
 & zxing & 5\\
 & okio & 5\\
 & fastjson & 5\\
\hline \hline
Total & 19 projects & 250 classes\\
\hline
\end{tabular}}
\label{table:2}
\end{table}
\subsection{SBST Tool}
In order to generate test cases for the classes under our experiment, we use EvoSuite~\cite{evosuite,test-suite}, which is an open-source tool that has been employed in many tuning studies~\cite{andrea,khodam}, in the past. For a given Java class bytecode, EvoSuite can generate a test suite with different objectives, such as maximizing branch coverage, using evolutionary algorithms~\cite{benchmark}. In this experiment, we made use of the most recent version of EvoSuite at the time, which was 1.0.6.

There are different variants of Genetic Algorithms implemented in the EvoSuite itself. In this experiment, for the sake of simplicity, we use the Monotonic GA implementation, which is the default one.

The GA setting employed in EvoSuite is configurable. If no parameter is specified for EvoSuite, the default setup which is selected by ``best practices'' will be used~\cite{evosuite}. Experiments have shown that ``EvoSuite default configuration'' can result in good coverage~\cite{andrea}.
\subsection{Search Space}
\label{search_space}
In hyper-parameter tuning, designing the search space, i.e. the parameters to tune and their possible values, can significantly affect the results. For a genetic algorithm, many parameters can be tuned~\cite{controlGA}. However, in order to be fair when compared with the related work, we used the same search space, i.e. same parameters and same values, as was previously used in EvoSuite tuning studies~\cite{andrea,khodam}.
Therefore, in this study, we iterate the values for five hyper-parameters of GA i.e. Crossover Rate, Population Size, Elitism Rate, Selection Function, and Parent Replacement Check. 

The range of values for these parameters is as follows:
\begin{itemize}
    \item \textit{Crossover rate:} {0, 0.2, 0.5, \textbf{0.75}, 0.8, 1} (6 values)
    \item \textit{Population size:} {4, 10, \textbf{50}, 100, 200} (5 values)
    \item \textit{Elitism rate:} {0\%, \textbf{1\%}, 10\%, 50\%} (4 values)
    \item \textit{Selection function:} roulette wheel function, tournament with two different sizes: 2 or 10, and \textbf{rank selection} with two biases: 1.2 or 1.7 (5 values)
    \item \textit{Parent replacement check:} \textbf{true} or false (2 values)
\end{itemize}

The default values of parameters in the version of EvoSuite that we used (i.e. 1.0.6) are highlighted in \textbf{bold}.

Hence, the search space that we are exploring here can generate $6 \times 5 \times 4 \times 5 \times 2 = 1,200$ distinct configurations for EvoSuite hyper-parameters.

\subsection{Experiment Procedure}
In the following section, the experimental procedure for our five research questions is explained:
\subsubsection{RQ1 (Class Prioritization)}
For all 250 classes under study, the four categories of static code metrics explained in section \ref{background} are calculated. In total, there are 105 different features extracted from 4 different categories.
Then, the search space defined in Section \ref{search_space} containing 1,200 configurations is created using different combinations of values for 5 hyper-parameters of genetic algorithms.

Then, all 250 classes under study are evaluated with EvoSuite with a fixed maximum budget of 2 minutes (that is twice as the EvoSuite's default budget to be consistent with other EvoSuite tuning studies~\cite{andrea, khodam}) per class, per configuration. These experiments are repeated 10 times, with different random seeds, to address the randomness nature of the algorithms. In total, there are $250 \times 1,200 \times 10$ experiments each taking 2 minutes to be evaluated by EvoSuite. While the execution time for all these experiments could take about 11.5 years on a single-core computer, we managed to run them in parallel~\cite{parallel} to effectively reduce the run time. All results and codes to conduct the same experiments are provided in the replication package\footnote{https://github.com/icsme2020author/Class-Prioritization-for-Tuning \label{package}}.

As a result, considering all of the configurations and repetitions, we had 12,000 EvoSuite evaluations, per class. We analyzed these observations carefully and calculated several statistical metrics for them such as their variances, the maximum values, and the median values. We then calculated our defined metrics mentioned in section \ref{method} to measure their ``Tuning Gain''.

To verify if we can prioritize classes properly, we split our data into train and test sets, keeping 40\% of classes (i.e. 100 classes) for test and we repeat the train-test splitting 100 times with different seeds and report statistics about the resulting distributions. 

{\it RQ1.1 design:}\\
In RQ1.1, we train a Random Forest Regression (RFR) with 200 trees (with max length = 5) on the described training sets to predict the ``Tuning Gain'' of classes on the test set. The classes are then sorted based on their predicted ``Tuning Gain'', in descending order. The goal of this ranking is to let the tuning method use the top classes, per given tuning budget. 

An important decision in our class-level prioritization is the selection size (number of classes that we want to allocate the tuning budget to).  This number defines the tuning budget per class (budget per class = total budget/ selection size). In RQ1.1, we first assess the effectiveness of our approach for any given selection size, by reporting the cumulative ``Tuning Gain'' of classes over different selection sizes (number of classes). We then summarize these measurements by calculating the Area Under Curve (AUC), when x-Axis is the selection size and y-Axis is the cumulative ``Tuning Gain''.

We compare the AUC of Random Forest Regression with a Random Ordering, a simple Linear Regression model, and the Optimal Ordering within our exhaustively evaluated hyper-parameters search space. 

In general, the fewer the number of classes, the more tuning budget per class. So the goal is to avoid spending the limited tuning budget on classes with less ``Tuning Gain''. Given that previous studies~\cite{khodam} have shown that around 80\% of classes in a project are insensitive to the configuration values, in the context of SBTG tuning, in RQ1.1, we also find the minimum selection size where the budget can be best spent. That is when the cumulative ``Tuning Gain'' curve becomes (almost) flat as the number of classes increases. In other words, tuning more classes is not adding any benefit.

As we will see in the results section, our experiment will partially confirm the 20-80\% rule of previous studies. Therefore, in the RQ2, we set the number of classes to 20\%. However, we will also study the effect of varying this cut-off value, in the effectiveness of our approach in RQ3.

In RQ1.1, we also look at the selection size in more detail by reporting the median of Normalized Cumulative Gains (NCG)~\cite{NCG} over the 100 train-test splits data, per prioritization method. The NCG is calculated as the Cumulative ``Tuning Gain'' of a method divided by the optimal Cumulative ``Tuning Gain''.

{\it RQ1.2 design:}\\
In RQ1.1, we used the entire collection of 105 features (static code metrics) in the regression models. In RQ1.2, we argue that most likely, there is a high correlation between these metrics and a smaller subset of features could be potentially as good as the entire set. Thus, in practice, we only need to collect a subset of metrics, per class. 

We utilize a feature selection method called Recursive Feature Elimination (RFE)~\cite{RFE}, which accepts a regression model and a dataset. It starts by building a model using all features and then eliminates the feature with the least predictive power (lowest coefficient). Next, RFE builds a new model, this time only with the remaining features. RFE repeats these steps until reaching the desired features set size. Given that Random Forest Regression (RFR) turns out to be the best model in RQ1.1 (see the results section for more details), we only use RFR within RFE. We also fix the selection sizes as 20, as discussed. In addition, we also report the median NCGs, over the 100 train-test splits.

The goals of this sub-RQ are (1) to see if we can achieve a cumulative gain as high as RQ1.1 (using all features) using only a small subset of features and (2) if so, which features are the more predictive ones.

\subsubsection{RQ2 (Tuning)}

To answer this question, we compare the coverage of all classes using 8 different tuning strategies, as follows:

\begin{itemize}
\item {\bf Default:} {The EvoSuite default configuration. Here we do not have any tuning. To be fair, for any given tuning budget, we let EvoSuite use more time (not just 2 minutes anymore), proportionally per class, to fill up the same time as other methods that have the tuning overhead.}

\item {\bf Global Meta-GA (Glob\_MG):} {This is the traditional tuning. We use the Meta-GA method but we could use any other tuning method as well. The key is how to allocate the tuning budget. The Global Meta-GA is not at the class-level. In other words, the whole project (all classes \textit{together}) are tuned at the same time and will have the same configuration (not tailored per class).}

\item {\bf Random Subset - Random Search (Rnd\_RS):} {This is a class-level version of random search method. That means each class is tuned individually (using a random search across the search space of hyper-parameters), but we do not allocate the budget equally to all classes of the project. Instead, we only select a subset of classes to be tuned and the rest will use the default parameter values. However, the classes are selected randomly. Subset size can vary, in all four class-level tuning approaches described here, depending on the classes and will be investigated in RQ3.}

\item {\bf Random Subset - Meta-GA (Rnd\_MG):} {This is a class-level version of the Meta-GA method. That means each class is tuned individually, but this time we do not allocate the budget equally to all classes of the project. Instead, we only select a subset of classes to be tuned and the rest will use the default parameter values. However, the classes are selected randomly (similar to Rnd\_RS).}

\item {\bf Random Subset -  Differential Evolution (Rnd\_DE):} {This method is exactly like Rnd\_MG except it uses a Differential Evolution algorithm rather than Meta-GA.}

\item {\bf Prioritized Subset - Random Search (Pri\_RS):} {Like Rnd\_RS, this is also a class-level tuning (individual classes are tuned, independently) and the budget is allocated to only a subset of classes (the rest use default parameter values). The same as Rnd\_RS it uses a random search across the search space of hyper-parameters for tuning. Unlike Rnd\_RS, however, this approach uses the prioritized list of classes from RQ1 and select a subset from top of the list (with the highest predicted ``Tuning Gain''). }

\item {\bf Prioritized Subset - Meta-GA (Pri\_MG):} {Like Rnd\_MG, this is also a class-level Meta-GA (individual classes are tuned, independently) and the budget is allocated to only a subset of classes (the rest use default parameter values). Unlike Rnd\_MG, however, this approach uses the prioritized list of classes from RQ1 and select a subset from top of the list (with the highest predicted ``Tuning Gain'').}

\item {\bf Prioritized Subset -  Differential Evolution (Pri\_DE):} {This method is exactly the same as Pri\_MG except it uses Hill a Differential Evolution algorithm rather than Meta-GA.}

\end{itemize}

The specifications of the mentioned tuning strategies is summarized in Table \ref{table:3}.

\begin{table*}[h!]
\centering
\caption{Summary of the six tuning strategies.}
\resizebox{\textwidth}{!}{\begin{tabular}{|lllll|} 
\hline
 \rowcolor{lightgray} Strategy & Tuning Method & Tuning Granularity & Budget Allocation & Selection Method \\ [0.5ex] 
 \hline
 Default & No Tuning & -- & --  & -- \\
 Global Meta-GA & Meta-GA & All Classes & All Classes & -- \\
 Random Subset Random Search & Random Search & Individual Classes &  Subset of Classes & Random \\
 Prioritized Subset Random Search & Random Search & Individual Classes & Subset of Classes & Prioritized\\
 Random Subset Differential Evolution & Differential Evolution & Individual Classes &  Subset of Classes & Random \\
 Prioritized Subset Differential Evolution & Differential Evolution & Individual Classes &  Subset of Classes & Prioritized \\
 Random Subset Meta-GA & Meta-GA & Individual Classes &  Subset of Classes & Random \\
 Prioritized Subset Meta-GA & Meta-GA & Individual Classes &  Subset of Classes & Prioritized \\
 \hline
\end{tabular}}
\label{table:3}
\end{table*}

{\it RQ2.1 and RQ2.2 design:}\\
In RQ2.1 and RQ2.2, we will compare six tuning methods in terms of their effectiveness with Default and Glob\_MG, as baselines.

Therefore, as discussed in RQ1.1 design, the number of classes we select in our six tuning methods would be 20\% of 100 test set classes i.e. 20 classes. The tuning subset that these tuning methods look into in one version comes from the prioritized list of classes and in the other version comes from random selection.

Two of the tuning strategies (Pri\_RS and Rnd\_RS) use Random Search. In Random Search, we first calculate the budget that we can allocate per class. This is dependent to the budget that we have for tuning and the number of classes that we want to tune. The budget allocation is not weighted and all classes are given budget equally. Then, we select and evaluate random configurations from the search space as many as the budget allows for each class. The configuration that yields highest extra covered branches compared to the default configuration is the tuned configuration for that given class. Ultimately, for calculating the extra covered branches of the class subset, we add up the extra covered branches of its classes. This whole process is repeated 25 times and the median value for extra covered branches is used in the results to avoid the effect of randomness.

The Pri\_MG and Rnd\_MG tuning methods and the Glob\_MG baseline use Meta-GA, which is explained in section \ref{background}. As the upper method, Meta-GA itself is a form of GA, and we should set its hyper-parameters, as well. In order to spend a reasonable time on tuning, we limited the size of the initial population of our Meta-GA to 6 configurations. The crossover rate is set to 0.5 and the mutation rate is set to 0.1 for this experiment. These values are similar to defaults GA parameters in EvoSuite~\cite{mutation}. 

The remaining two methods (Pri\_DE and Rnd\_DE) replace Meta\_GA with Differential Evolution in Pri\_MG and Rnd\_MG and keep all other parameters unchanged. In our implementation, we kept classical values for crossover probability (CR) and differential weight (F) i.e. 0.8 and 0.9, respectively~\cite{differential_tuning}.

We repeat this experiment with different tuning budgets. We start as low as one hour and increase the budget with increments of an hour up to 24 hours. To put these numbers in context, as we have 100 classes in the test set for tuning, each extra hour is translated to 36 extra seconds of search budget per class (60 minutes/100 Class = 36 seconds), for the Default configuration, which does not use any tuning.

In order to compare the results of tuning over all budgets (from 1H to 24H), we report both (1) AUC of each graph and (2) the average of extra covered branches over all budgets to make the numbers more understandable. In order to find the average value of extra covered branches, we calculate the area under curve and divide it by the range of budget which is 24 Hours in our case.

\begin{displaymath}
Average Extra Covered Branches = \frac{AUC}{Budget Range}
\end{displaymath}

In addition, we test the statistical significance of our comparisons each time (over the 25 repetitions and 100 train-test splits) by running a Mann–Whitney U test on the results to confirm that the differences between medians are statistically significant.

{\it RQ2.3 (Cost Analysis) design:}

In RQ2.3, we focus on the tuning cost and look into two aspects of the cost: (1) whether our tuning method is particularly better for some tuning budgets (low vs high) and (2) whether our tuning method overhead is justifiable in practice. 

The challenge regarding low tuning budgets (less than 4 hours here) is that for techniques that allocate budget to all classes (Tuning Subset size is 100\%), we can not even finish one complete iteration of Meta-GA or DE. To solve this issue, we had to allocate the resources to a random subset of classes. For instance, if our budget is set to 1 hour, we have to select 5 classes out of 100 classes to be able to evaluate them for one iteration of Meta-GA in Global Meta-GA.  Therefore, we did not go for less than an hour as our minimum budget (essentially most methods would become random at very low budgets). We also stopped at 24H as our maximum, in this RQ, to emulate a one-day tuning job. We investigate the trends for higher budgets in RQ5.

To answer (2), we have a discussion on extra coverage vs. tuning budget of our approach and Default and also explain which tuning methods can benefit more from parallel computing.

\subsubsection{RQ3 (Tuning Subset Comparison)}

RQ3 investigates the trade-off between shallow tuning of more classes and deep tuning a few classes. We will examine this by changing the size of the Prioritized Subset of classes under study.

This RQ is firstly motivated by the results of RQ1.1., where the 20\% and 60\% cut-offs both seem reasonable, and secondly by the results of RQ2, where the extra coverage of Pri\_MG and Pri\_DE kept increasing, until the tuning budget was high enough to tune all the classes in the subset for one Meta-GA or DE iteration. However, after that point, adding more budget did not help significantly in covering more branches. 

Therefore, we assume that the 20\% cut-off value might not be the optimal. Thus, we repeat the same instruction for our four class-level approaches, for subset size of 60\% and 100\% of the classes, in the test set, in RQ3.

\subsubsection{RQ4 (Search Space Size)}
In this RQ, we want to see if reducing the search space size can help improving the search's effectiveness. The motivation is if we exclude ``insignificant'' hyper-parameters (those that won't matter much in the search-space landscape), we can explore more of the search-space with the same budget.

In order to eliminate insignificant hyper-parameters from the search space, for each class in the training set, we analyze the fitness landscape of each class to find the hyper-parameters along which the fitness landscape is ``neutral'', i.e. mostly flat and has no change in fitness value~\cite{Rugged}.
As a simple heuristic, we use non-linear regression models and use the coefficient of each hyper-parameter as a measure of how neutral the landscape is along that hyper-parameter. The less the coefficient of a hyper-parameter, the more neutral fitness landscape is along that axis.

We build a regression model using Random Forest Regressor (with 200 trees and max-depth of 5) in which the data-points are represented by their configurations and the resulted coverage values, per configuration. In other words, the independent variables are the five hyper-parameters of a configuration, and the dependent variable is ``Covered Branches'', for that given class and configuration.
Each model can predict the covered branches for a given class at any configuration. We will calculate the vector of feature importance for each model that translates into the importance of each hyper-parameter in each class. Feature importance is a value between 0 and 1, and they add up to 1.

We then use RFE (already explained in RQ1.2 Design) to eliminate these insignificant hyper-parameters (hyper-parameters that appear as the least important hyper-parameter in most classes) one-by-one to reduce the size of search space. The default value is used for a hyper-parameter when it is eliminated from the search space.

Given that we have 5 hyper-parameters, we create 3 reduced versions of search space after eliminating one hyper-parameter in each step of RFE. Table \ref{tab:rq4} lists the number of hyper-parameters (in the results section we will report the exact hyper-parameters discarded at each step) and the total size of each search space.

\begin{table}[h!]
    \centering
        \begin{tabular}{|ccc|}
    \hline
        \rowcolor{lightgray} Search Space Size & No. of HPs & Search Space Size\\
        \hline
        Full & 5 & 1200 \\
        Large & 4 & 600 \\
        Medium & 3 & 100 \\
        Small & 2 & 20 \\
        \hline
    \end{tabular}
    \caption{Search Spaces after Hyper-parameters Elimination}
    \label{tab:rq4}
\end{table}

We then replicate RQ3 with the large, medium, and small search spaces, to see if reducing the search space can help the search methods to find the tuned configuration faster.

\subsubsection{RQ5 (Allocating high tuning budgets)}
In this research question, we want to see if we can improve the performance of search methods by allocating longer time for tuning, even if it is impractical. The goal of this question is to find out the difference between search methods in higher budgets, where more complex methods are supposed to work better, at least in theory. 

Based on the results from previous RQs, we use 60\% as the cut-off for class subset size and compare Pri\_MG and Pri\_RS with budgets of 100H to 500H, increments of 100H. 
Additionally, based on the results in RQ4, we repeat the experiment with a smaller search space. We want to see if making search space smaller and allocating higher budgets, at the same time, can make any changes to the previous observations.

\subsection{Execution Environment and Implementation Details}
\label{sec:env}

Our implementation for pre-processing data, building the regression models, and calculating metrics like AUC leveraged the scikit-learn library~\cite{scikit-learn} in python. The MetaGA method uses the implementation of GA in EvoSuite~\cite{evosuite} in Java. The DE method uses the same GA implementation in EvoSuite with some minor changes to implement DE.

We designed our execution scripts to run small tasks (calculating coverage for a class given a configuration for 10 times) in the cloud. Therefore, we were able to run these small tasks in parallel over 3 clusters (the name of clusters are redacted due to double-blind requirements) each with hundreds of nodes (32 to 48 CPU cores per node). In this study, $250 \times 1,200 \times 10$ (=3 millions) tasks were conducted with 2 minutes EvoSuite search budget, and $250 \times 10 \times 24$ (=60,000) tasks with on average 8 minutes budget (when using more budget for Default setup in RQ2.2). The whole experiment would take around 12 years on a single core setting, however, with the help of parallel computing and large nodes, it took a few months to conduct this research (including all the queuing and interruptions).

\subsection{Results}
In this section, we summarize results and discuss their implications. 
\subsubsection{RQ1 Results: How  accurately  static  source  code  metrics  can predict the cost-effectiveness of SBTG tuning, per class?}

There are two sub-research questions in RQ1:

{\it RQ1.1: How accurately static source code metrics of a class can predict its ``Tuning Gain'', using a random forest regression?}

\begin{figure}[h!]
\centering
\includegraphics[width=\linewidth]{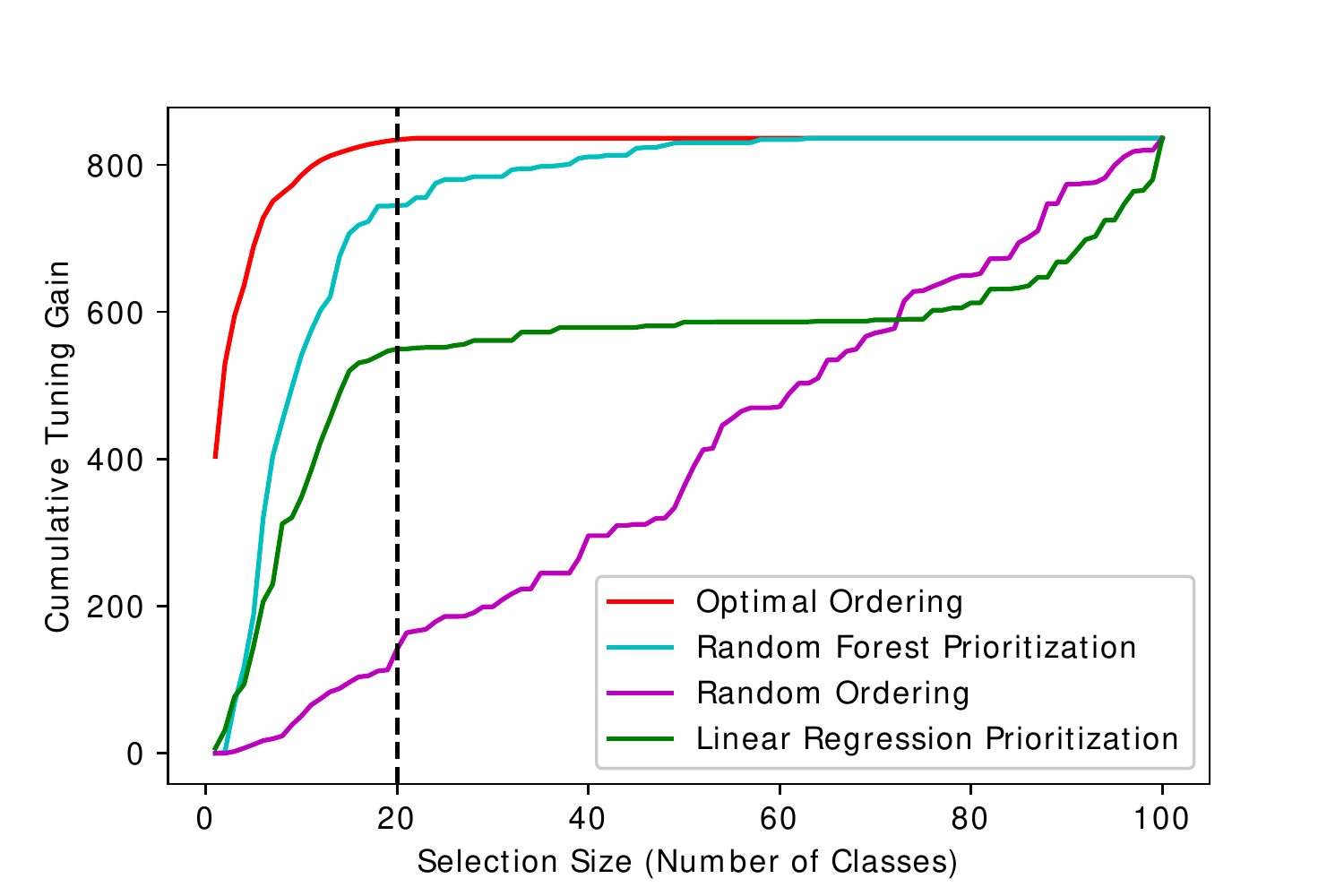}
\caption{Cumulative sum of Tuning Gain over different selection sizes (number of classes)}
\label{fig:rq11}
\end{figure}

To answer this sub-RQ, as discussed, we look at the prediction results in terms of Cumulative ``Tuning Gain'' as well as AUC of Cumulative ``Tuning Gain'' over all selection sizes. Figure \ref{fig:rq11} represents Cumulative ``Tuning Gain'' for class prioritization using Random Forest and Linear Regression, as well as Random and Optimal ordering of classes. The X-axis in the figure represents the selection size which is the number of classes to select and the Y-axis shows the median Cumulative ``Tuning Gain'', when the prediction is repeated 100 times (with different seeds), per selection size (number of classes) and prioritization method.

Now looking at different prioritization methods, we see that, as expected, a random ordering acts quite poorly and can not prioritize the best classes over others. The two regression models, however, are way better and with around the first 20\% of classes they capture most of the cumulative ``Tuning Gain''. The Random Forest approach is particularly superior, since Linear Regression wrongly predicts some of the worthy classes (with high ``Tuning Gain'') at the bottom of the list (see the jump in ``Tuning Gain'' for Linear Regression, from 80 to 100 classes). In fact, Random Forest Regression performs the same as the Optimal ranking from 60\% selection size onward.

In order to summarize and compare these methods over all selection sizes, we use the ``AUC ratio'' metric, where the AUC of each of these curves is divided by the AUC of the ``Optimal Ordering'' (the red curve).

As it is shown in Figure \ref{fig:rq11}, and is confirmed by the AUC ratio values in Table \ref{tab:AUC}; regression models are rather successful in prioritizing the classes across all selection sizes. However, the data also show that Random Forest Regression (the blue curve) outperforms Linear regression by 36\% ($(892/658)-1.0$), in terms of the AUC ratio. 

Another observation based on Figure \ref{fig:rq11} is that the cumulative ``Tuning Gain'' reaches a maximum with only around 20\% of the classes. In other words, ranking those 20\% upfront classes is quite crucial. This is in line with the previous finding~\cite{khodam} that around 80\% of classes are insensitive to tuning.  

To look at the cumulative gain at the 20\% selection size in more detail, Table \ref{tab:AUC} reports the median of the Normalized Cumulative Gains (NCG) over 100 experiments with different train-test splits, all using 20\% selection size. As we can see, the results confirm all the above findings. The Random Forest Regression, with 0.892 NCG, outperforms Linear Regression by 36\% ($0.892-0.658$) and outperforms Random selection by 70\% ($0.892-0.196$). 

Therefore, in the rest of this manuscript, (a) unless otherwise is stated, we use 20\% as the selection size whenever a method requires the class selection and (b) we use Random Forest Regression as our class prioritization method.

\begin{table}[htbp]
    \centering
    \caption{AUC Ratios and the Median NCGs of Different Prioritization Methods, in Figure \ref{fig:rq11}}
    \begin{tabular}{|c c c|}
    \hline
        \rowcolor{lightgray} Method & AUC Ratio & NCG Score (20 Classes) \\
        \hline
        Optimal Ordering& 1 & 1 \\ 
        Random Forest Regression & 0.921 & 0.892\\ 
        Linear Regression & 0.677 & 0.658\\
        Random Ordering & 0.483 & 0.196\\
        \hline
    \end{tabular}

    \label{tab:AUC}
\end{table}

{\it RQ1.2: Which static source code metrics are more powerful for ``Tuning Gain'' prediction?}

In this research question, we look at the Random Forest Regression results from two aspects: 1) can we eliminate some features without a significant loss in performance? and 2) what are the most predictive features in the regressions?

As discussed, we use the RFE feature selection method and fix the number of classes to 20. Table~\ref{tab:AUC2} reports the Normalized Cumulative Gain (NCG) of ``Tuning Gain'' when the Random Forest Regression only uses the top 2,5,10,20 and 40 features. It reports the median of the NCGs over 100 experiments with different train-test splits. 

The first observation is that with only 40 features we can achieve almost the same NCG (only 0.001\% loss) as the original model (with 105 features). That is almost 62\% saving in terms of static metric collection and computation. We can also observe that the losses in NCG when the model uses 20, 10, 5, and 2 features compared to the original 105 features are: 2\%, 5\%, 9\%, and 16\%, respectively. Therefore, we will use the top 40 features in the rest of this study, when using Random Forest Regression.

\begin{table}[h]
    \centering
    \caption{NCGs of Random Forest Regression with different number of features (selection size = 20 classes).}
    \begin{tabular}{|c c|}
    \hline
        \rowcolor{lightgray} Number of Features & NCG Score (20 Classes) \\
        \hline
        Top 2  & 0.747\\ 
        Top 5   & 0.813\\
        Top 10   & 0.848\\
        Top 20   & 0.875 \\
        Top 40   & 0.891 \\
        105 (No Selection)  & 0.892 \\
        \hline
    \end{tabular}
    \label{tab:AUC2}
\end{table}

With respect to the second aspect of RQ1.2 (the most predictive features), the most frequent set of top 2 features are: ``rfc'', ``else'' and ``Difficulty''(a Halstead complexity metric). 

In fact, the most recurring feature in the regression is RFC for all sets of features. RFC is a measure of complexity and it makes sense to have such a feature as a core metric to decide which classes are good for test generation tuning. The top 10 features when selecting 40 features using RFE are as follows: ``Difficulty'', ``rfc'', ``Abstract Class Count'', ``else'', ``anonymousClassesQty'', ``comparisonsQty'', ``Effort'', ``class'', ``do'' and ``stringLiteralsQty''. The full lists of most recurring features for all the feature selection sizes are provided in the replication package. 

\begin{tcolorbox}[drop shadow,enhanced,sharp corners,rounded corners=downhill,enlarge top initially by=5mm]
Random Forest Regression on Static Code Metrics accurately predicts the cost-effectiveness of SBTG tuning, per class, as measured by ``Tunability Gain'' (NCG of 89\% for class size 20 and AUC of 92\% over all selection sizes) and outperforms the alternatives.
\end{tcolorbox}

\subsubsection{RQ2 Results: How cost-effective is our class prioritization technique for SBTG tuning, compared to traditional tuning methods?}

\begin{figure}
\begin{subfigure}{\textwidth}
\centering
\includegraphics[width=\linewidth]{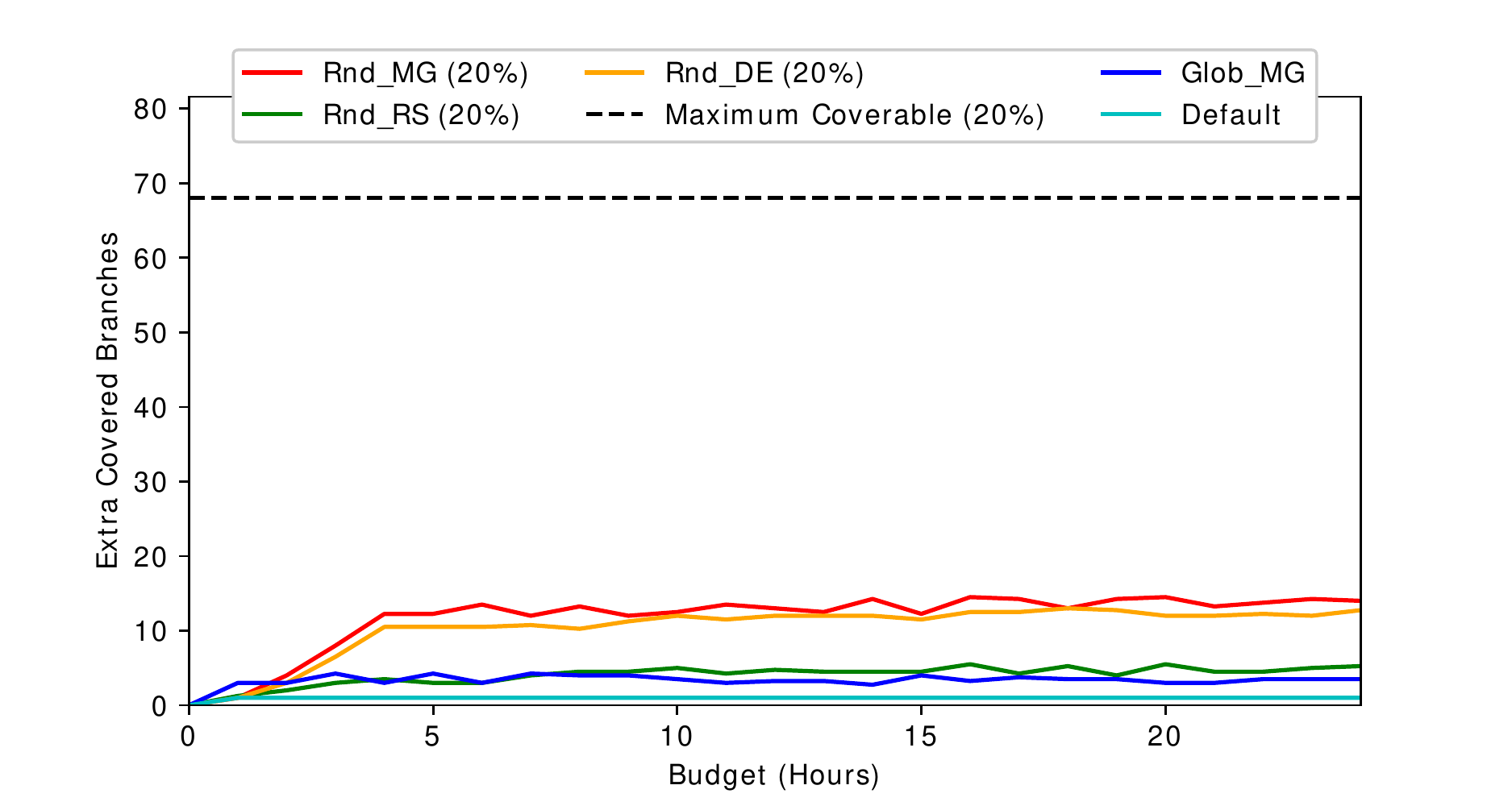}
\caption{Random Subset}
\label{fig2:sub1}
\end{subfigure}
\begin{subfigure}{\textwidth}
\centering
\includegraphics[width=\linewidth]{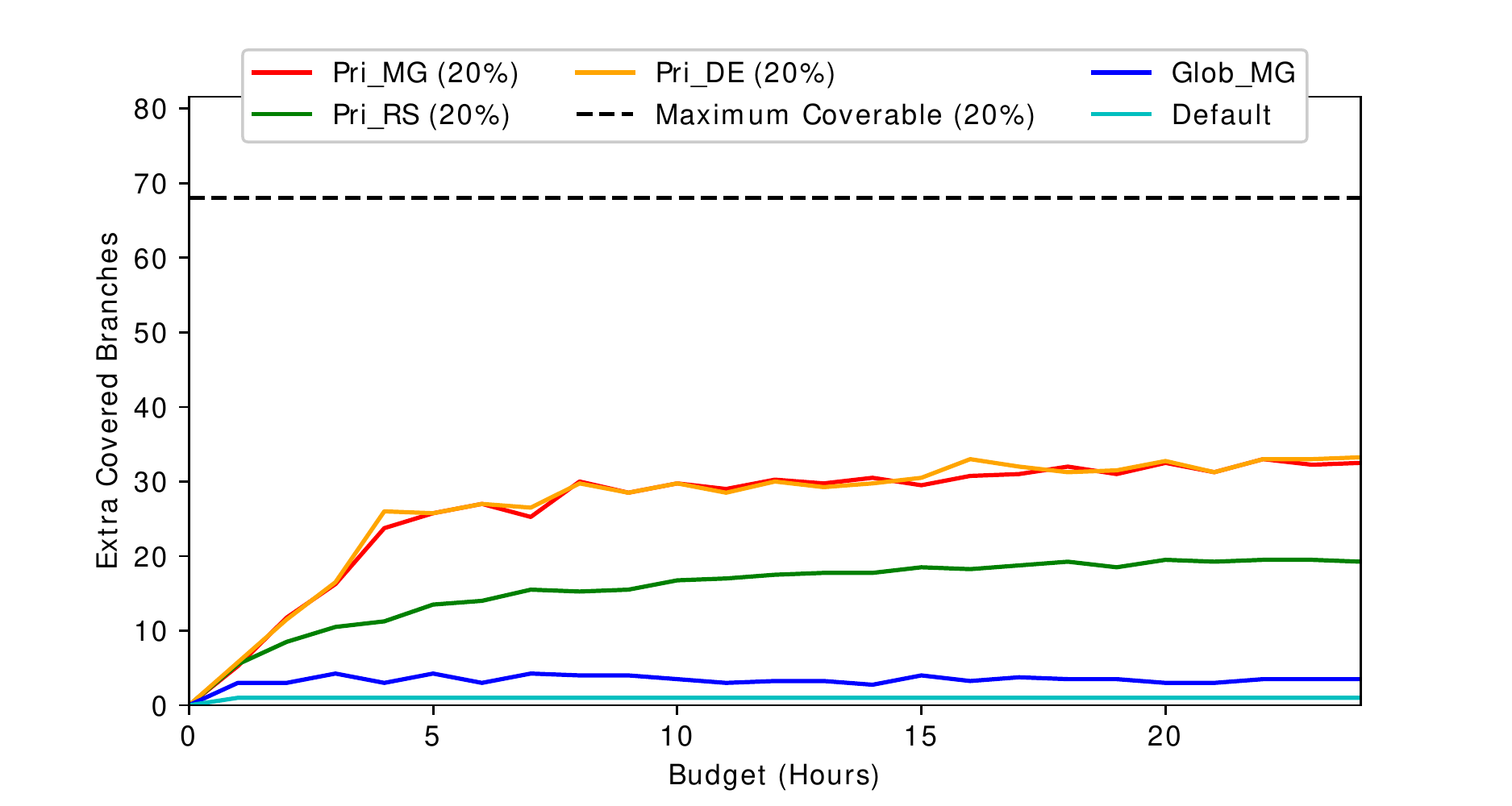}
\caption{Prioritized Subset}
\label{fig2:sub2}
\end{subfigure}

\caption{Extra Covered Branches vs. Tuning Budget (From 1H to 24 H increments of 1H. For class-level tuning the class selection size is 20\%.)}
\label{fig:rq2}
\end{figure}

Figure \ref{fig:rq2} reports the extra number of covered branches using generated test cases, when EvoSuite is tuned by one of the six class-level tuning methods compared to covered branches using, a global Meta-GA approach (our tuning baseline), and the default values (no tuning baseline), over the 100 classes in the test set. It also shows the maximum number of extra branch coverage for this set of classes. 

As discussed in the design section, the tuning budget is from 1 to 24 hours, the selection size is 20, and the results are the medians over 100 train-test split repetitions. Figure \ref{fig2:sub1} shows when a random subset of classes is selected and Figure \ref{fig2:sub2} shows the same experiment when prioritized classes are tuned. Looking at these Figures, we want to answer the question that how beneficial our approach is for different tuning budgets, compared to the alternatives.

{\it RQ2.1: How effective is our class prioritization method compared to a random ordering of classes, in terms of their achieved extra code coverage?}

To answer this question, we first focus on the results of tuning with and without prioritization for a given search method. Therefore, we compare Rnd\_RS with Pri\_RS as well as  Rnd\_DE with Pri\_DE and Rnd\_MG with Pri\_MG.
As we see in Figure \ref{fig:rq2}, for both search methods and for all search budgets sorting the classes with our proposed method discussed in RQ1 can improve the coverage significantly.

In order to quantify the improvements after prioritizing classes, we calculate the average extra coverage for each tuning method, dividing the AUC by the budget range (24 Hours), and reported the results in Table \ref{tab:AverageExtraCoverage}.

\begin{table}
    \centering
    \caption{Tuning results in terms of area under curve (AUC), average extra covered branches (ECD) and their p\_Values over 24 Hours of tuning with different methods, over 100 classes.}
    \begin{tabular}{|llcc|}
        \hline
        \rowcolor{lightgray} Tuning Method & AUC & Average ECB & p\_Values\\
        \hline
        Default & 23.52 & 0.98 &  8.4e-28 \\ 
        Global Meta-GA & 81.39 & 3.39 & 5.2e-23\\
        Random  Subset  Random Search & 147.60 &  6.15 & 1.56e-9\\
        Random  Subset  Differential Evolution & 250.63 & 10.44 & 5.5e-06\\
        Random  Subset  Meta-GA & 285.12 & 11.88 & 5.7e-06\\
        Prioritized Subset Random Search & 378.96 & 15.79 & 3.17e-08 \\
        Prioritized Subset Differential Evolution & 651.5 & 27.15 & 0.439\\
        Prioritized Subset Meta-GA & 717.61 &  29.90 & -\\
        \hline
    \end{tabular}
    \label{tab:AverageExtraCoverage}
\end{table}

Prioritizing classes for Meta\_GA can on average improve covered branches by 18.02 (29.90 - 11.88) extra, which is 152\% more than the Rnd\_MG. In addition, comparing Pri\_DE and Rnd\_DE, we see that Pri\_DE covers on average 16.71 (27.15 - 10.44) extra branches which is 160\% more than Rnd\_DE. Finally, using Random Search on Prioritized Subset covers 9.64 (15.79 - 6.15) more branches than a Random Search on a Random Subset, which is 157\% improvement. 
So overall, no matter what search method (MG, DE, or RS) to be used, the prioritization part of our algorithm is always beneficial and improves the effectiveness of tuning.

{\it RQ2.2:  How effective is the MetaGA search compared to alternative
search methods, within the hyper-parameters' search-space, in terms of their achieved extra code coverage?} 

In this question, we switch our focus to compare the search methods that were studied here i.e. Meta-GA, DE, and Random Search when class orders are the same. Looking at Figure \ref{fig:rq2}, for both (Pri\_MG, Pri\_DE, Pri\_RS) and (Rnd\_MG, Rnd\_DE, Rnd\_RS) pairs using either Meta-GA or DE as the search technique improves the final coverage.

Based on Table \ref{tab:AverageExtraCoverage}, for prioritized classes using Meta-GA can improve random search's coverage by 14.11 branches ($29.90 - 15.79$) equivalent to 89\% increase; and for random classes Meta-GA covers 5.73 branches ($11.88 - 6.15$) more than random search (93\% improvement). 

However, the differences between Meta\_GA and DE are negligible. The Mann–Whitney U tests suggest that the differences between Pri\_MG and all other techniques except Pri\_DE is statistically significant. Therefore, we suggest using either of Pri\_MG or Pri\_DE as the tuning method. In the rest of this paper, we use Pri\_MG, since its median results is slightly higher (although statistically insignificant) than Pri\_DE.

{\it RQ2.3: What is the cost of our SBTG tuning, in practice?}

Based on the results from RQ2.1 and RQ2.2, our proposed method of tuning (Pri\_MG: Prioritized Subset Meta-GA), which employs both prioritization of classes and Meta-GA, is the best method to tune classes, in terms of the extra covered branches. Results in Table \ref{tab:AverageExtraCoverage} and Figure \ref{fig:rq2} confirm this finding.

Now, to look at its cost-effectiveness, we compare Pri\_MG with two baselines: the Glob\_MG (as a tuning baseline; which is tuning all classes together using the same Meta\_GA) and Default (as a non-tuning baseline; where we use default parameter values in EvoSuite and the unused tuning budget will be given to EvoSuite to spend on the actual search for test cases).

We compare the techniques in two setups, when: (a) a low tuning budget is given, e.g., 4 Hours, and (b) when we have high budget for tuning, e.g., 24 Hours.

After 4H of tuning, we see that Pri\_MG covers 27.25 ($30.25 - 3.0$) and 29.25 ($30.25 - 1.0$) branches more than Glob\_MG and Default, which is around 10 times and 30 times improvement, respectively.

When a high budget (24H) is given, Pri\_MG outperforms Glob\_MG and Default methods by 32.25 ($35.75 - 3.5$) and 34.75 ($35.75 - 1.0$) branches, which is equivalent to 9.2 and 34.75 times improvement, respectively. This confirms that our approach is a cost-effective approach that improves the baselines significant both for low and high budgets, regardless of whether that extra budget is going to be used on tuning or on the actual test generation task.

So far, all we said is about how to spend a low/high tuning budget, to get maximum extra coverage. But whether the achieved extra coverage worth the tuning overhead is another question that is more difficult to answer. The improvements using our proposed method was around an extra 30 branches with 4H of tuning. Whether this is a practically cost-effective approach or not partly depends on the context of the application. If we are dealing with a more critical code-base, we conclude the 4H extra budget definitely worth the extra 30 branches, especially given that no other tuning method nor spending the 4H on EvoSuite directly can achieve this much improvement.

Finally, it is worth mentioning that our approach is highly parallelizable, which can significantly reduce the tuning costs. For instance, in our study a 4-hour Prioritized Subset Meta-GA on a single machine takes only 40 minutes with 6 concurrent threads each working on one class in the population, which is 83\% time-saving and much more affordable.

\begin{tcolorbox}[drop shadow,enhanced,sharp corners,rounded corners=downhill,enlarge top initially by=5mm]
Our approach, Pri\_MG (Prioritized Subset Meta-GA), which uses a novel class prioritization method to select a subset of classes and then applies a Meta-GA to tune those classes, covers extra 30 branches on 100 classes with 4H of tuning on a single machine (or 40 minutes in a basic cloud). This improves the state of the art tuning techniques by more than 10 times. We recommend using Pri\_MG for low budget tuning.
\end{tcolorbox}

\subsubsection{RQ3 Results: What is the effect of the number of classes selected for tuning on the overall effectiveness of our approach?}

\begin{figure}
\begin{subfigure}{0.5\textwidth}
\includegraphics[width=1.1\linewidth]{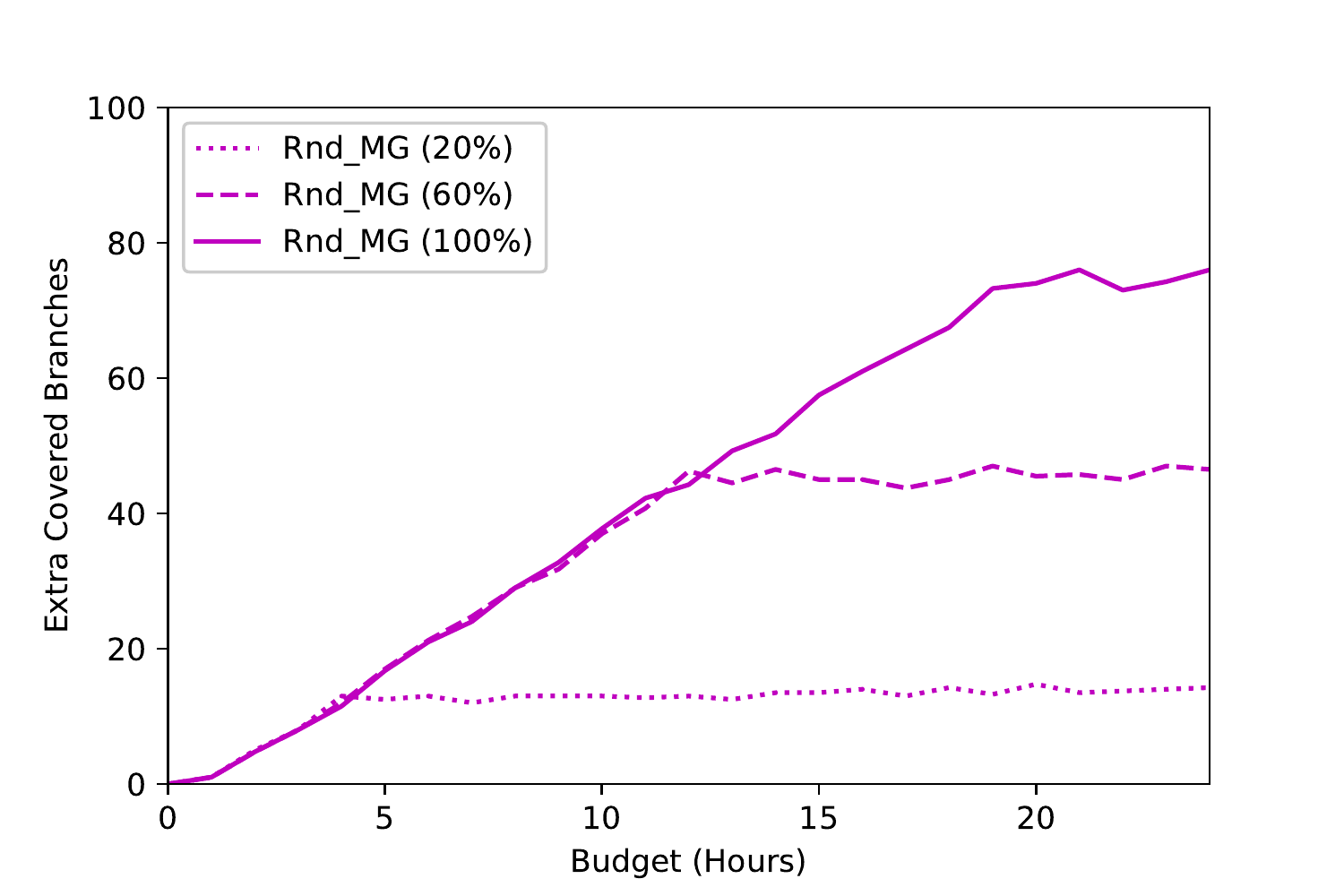}
\caption{Random Subset Meta-GA}
\label{fig:sub1}
\end{subfigure}
\begin{subfigure}{0.5\textwidth}
\includegraphics[width=1.1\linewidth]{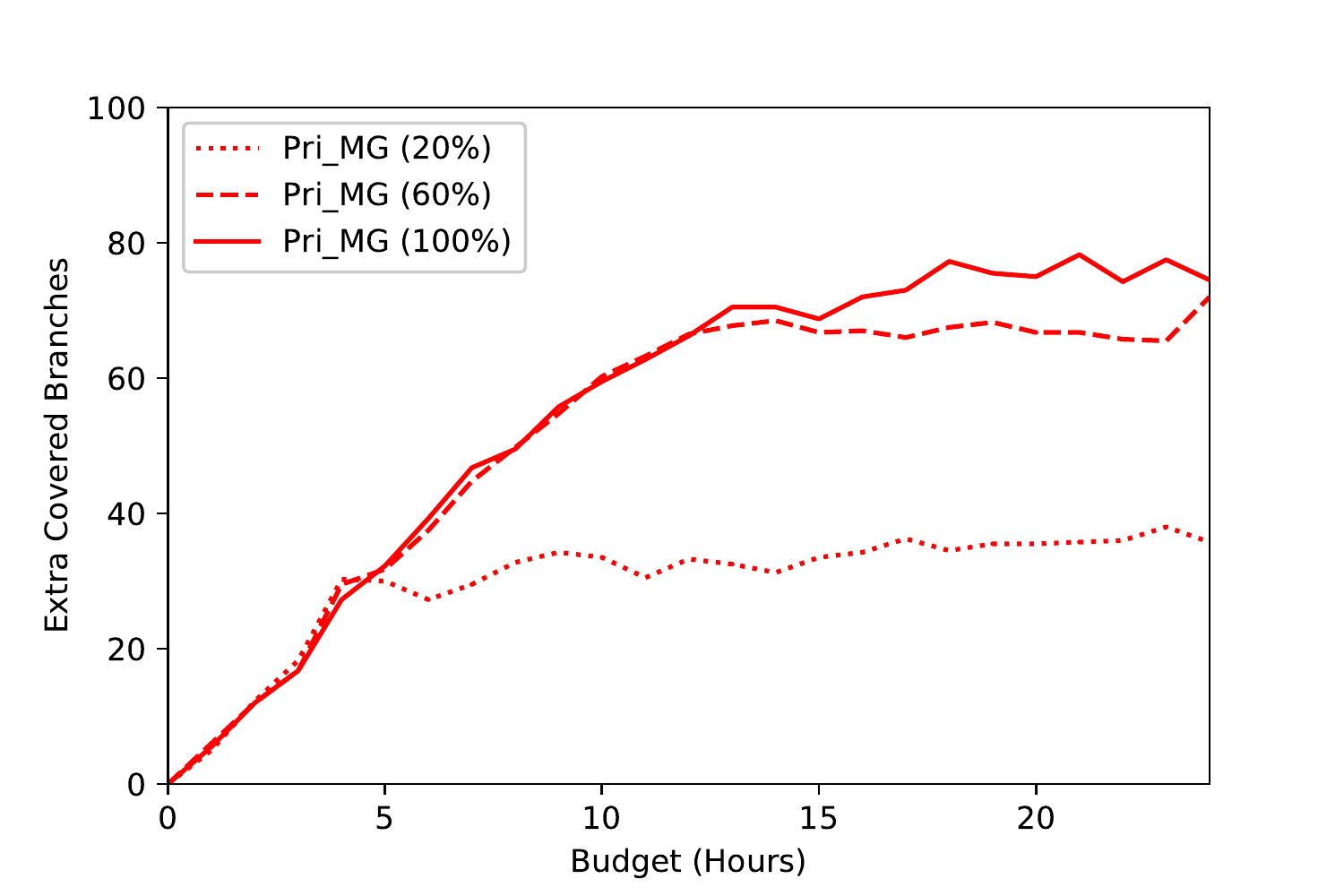}
\caption{Prioritized Subset Meta-GA}
\label{fig:sub2}
\end{subfigure}
\begin{subfigure}{0.5\textwidth}
\centering
\includegraphics[width=1.1\linewidth]{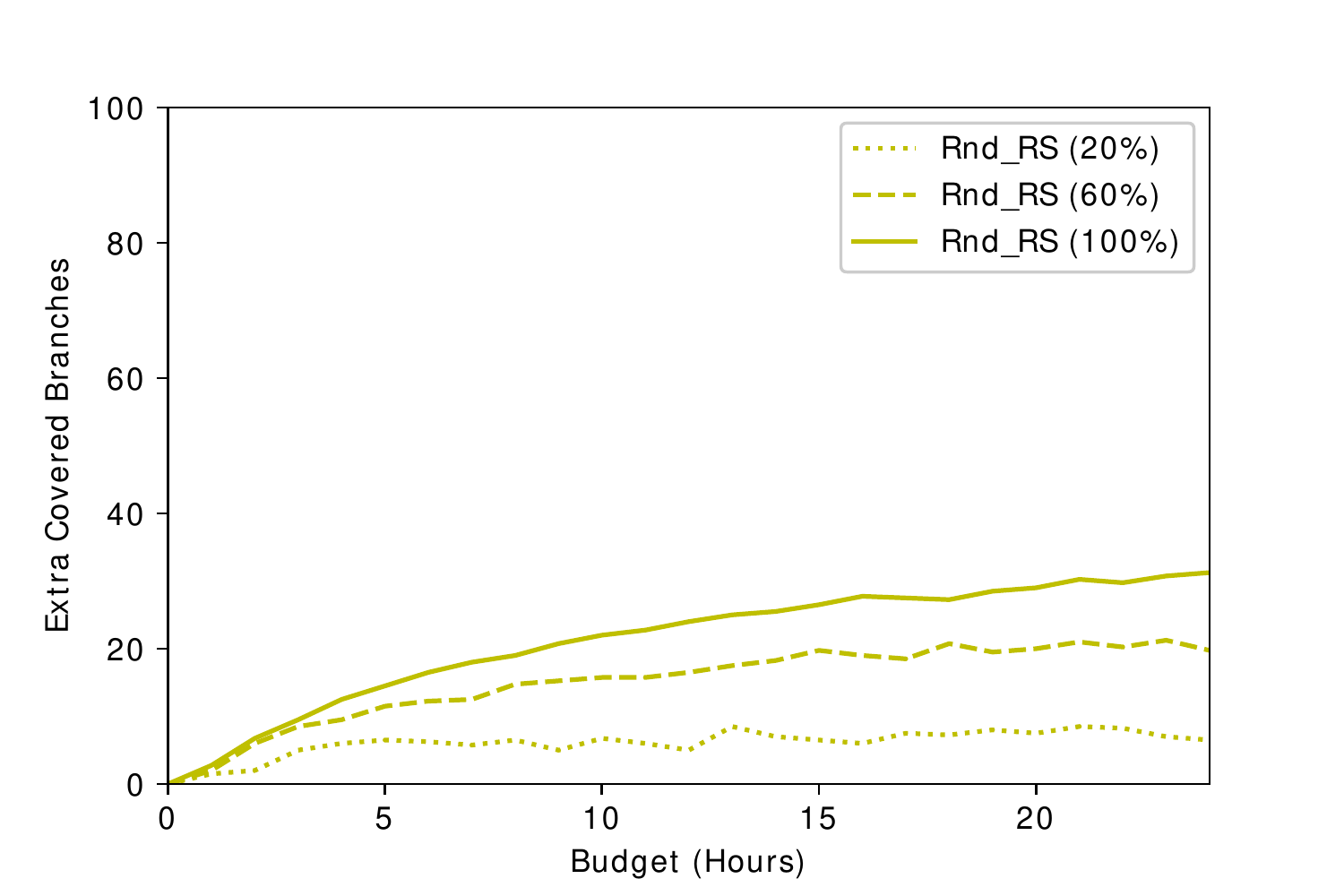}
\caption{Random Subset Random Search}
\label{fig:sub3}
\end{subfigure}
\begin{subfigure}{0.5\textwidth}
\centering
\includegraphics[width=1.1\linewidth]{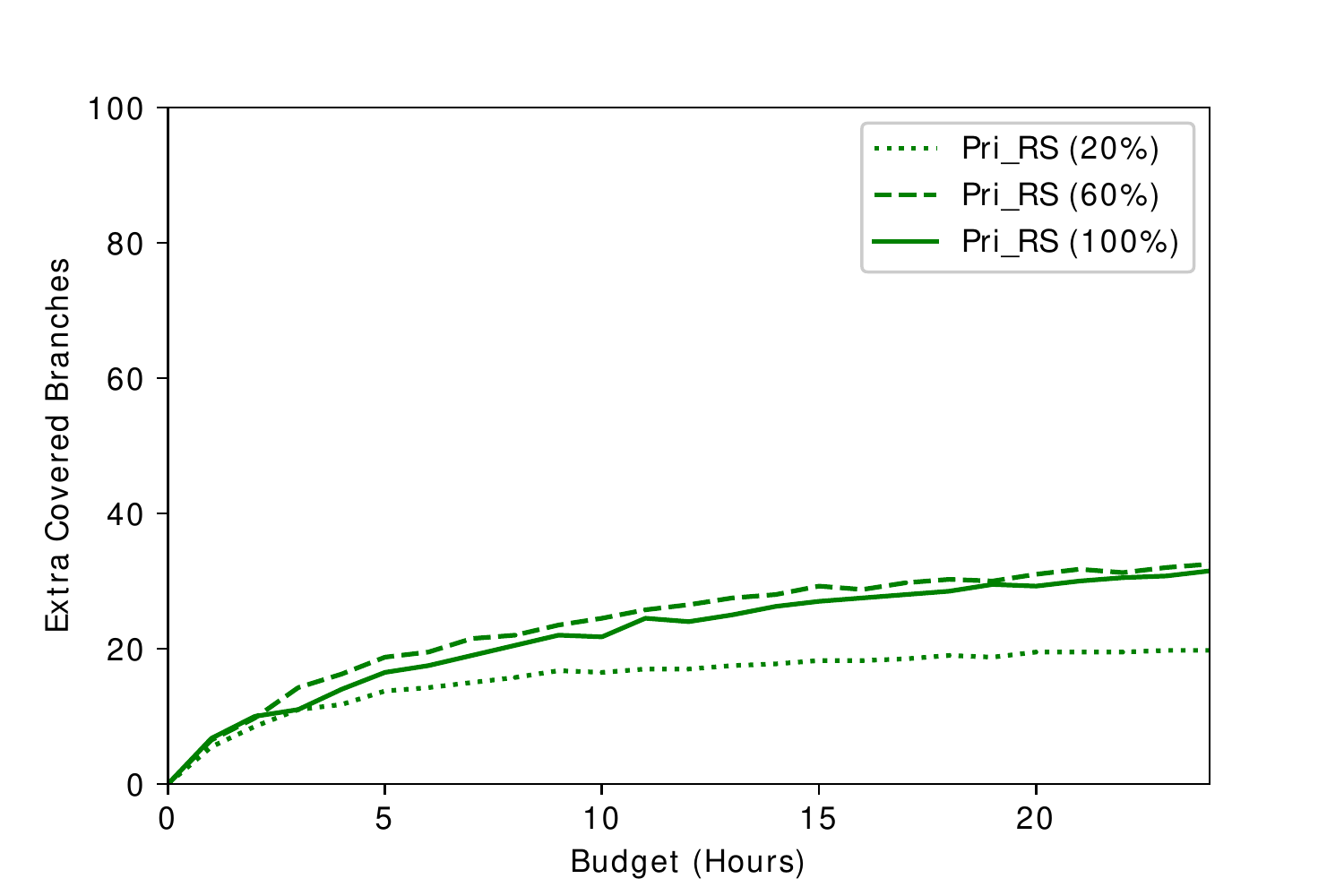}
\caption{Prioritized Subset Random Search}
\label{fig:sub4}
\end{subfigure}
\begin{subfigure}{0.5\textwidth}
\centering
\includegraphics[width=1.1\linewidth]{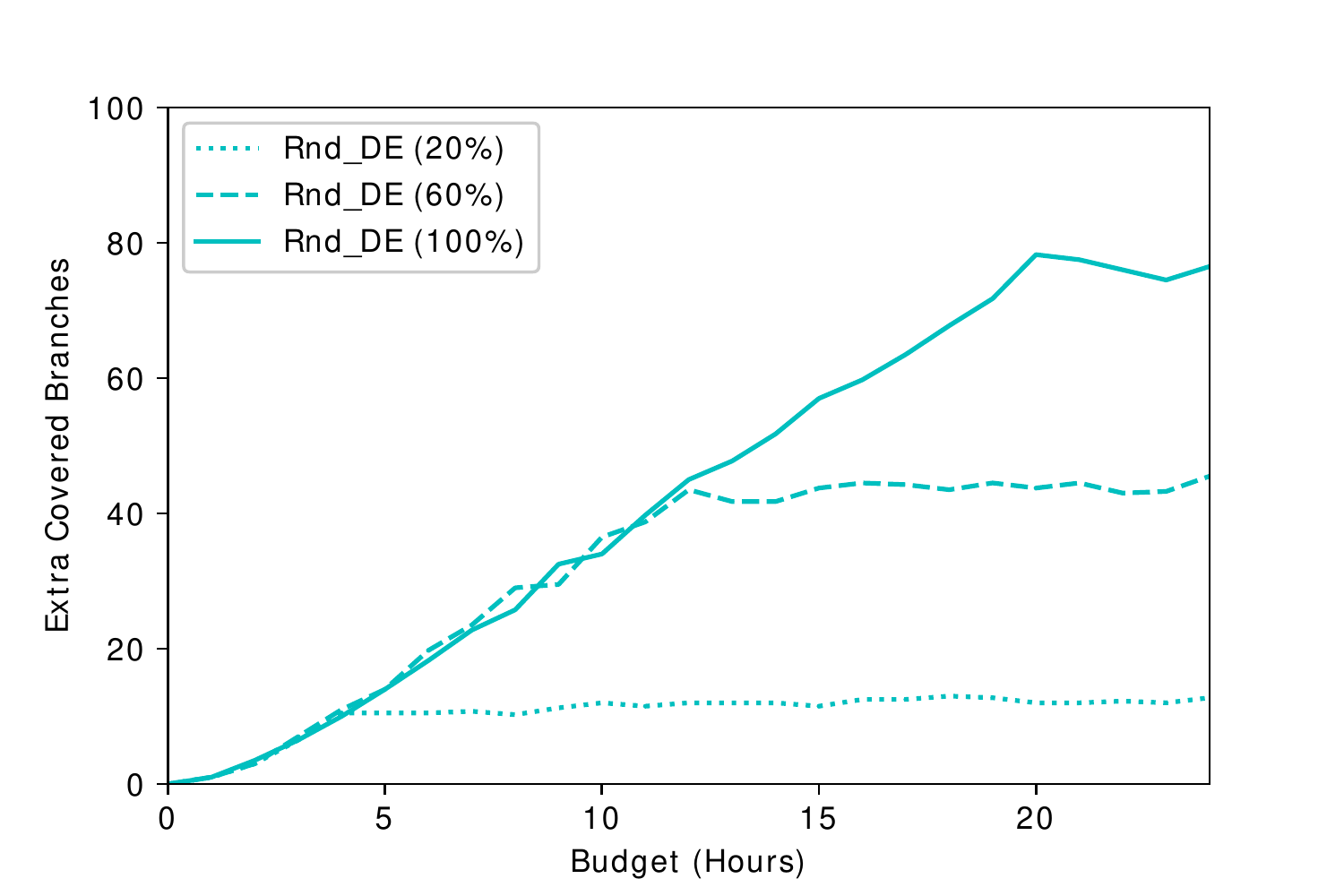}
\caption{Random Subset Differential Evolution}
\label{fig:sub5}
\end{subfigure}
\begin{subfigure}{0.5\textwidth}
\centering
\includegraphics[width=1.1\linewidth]{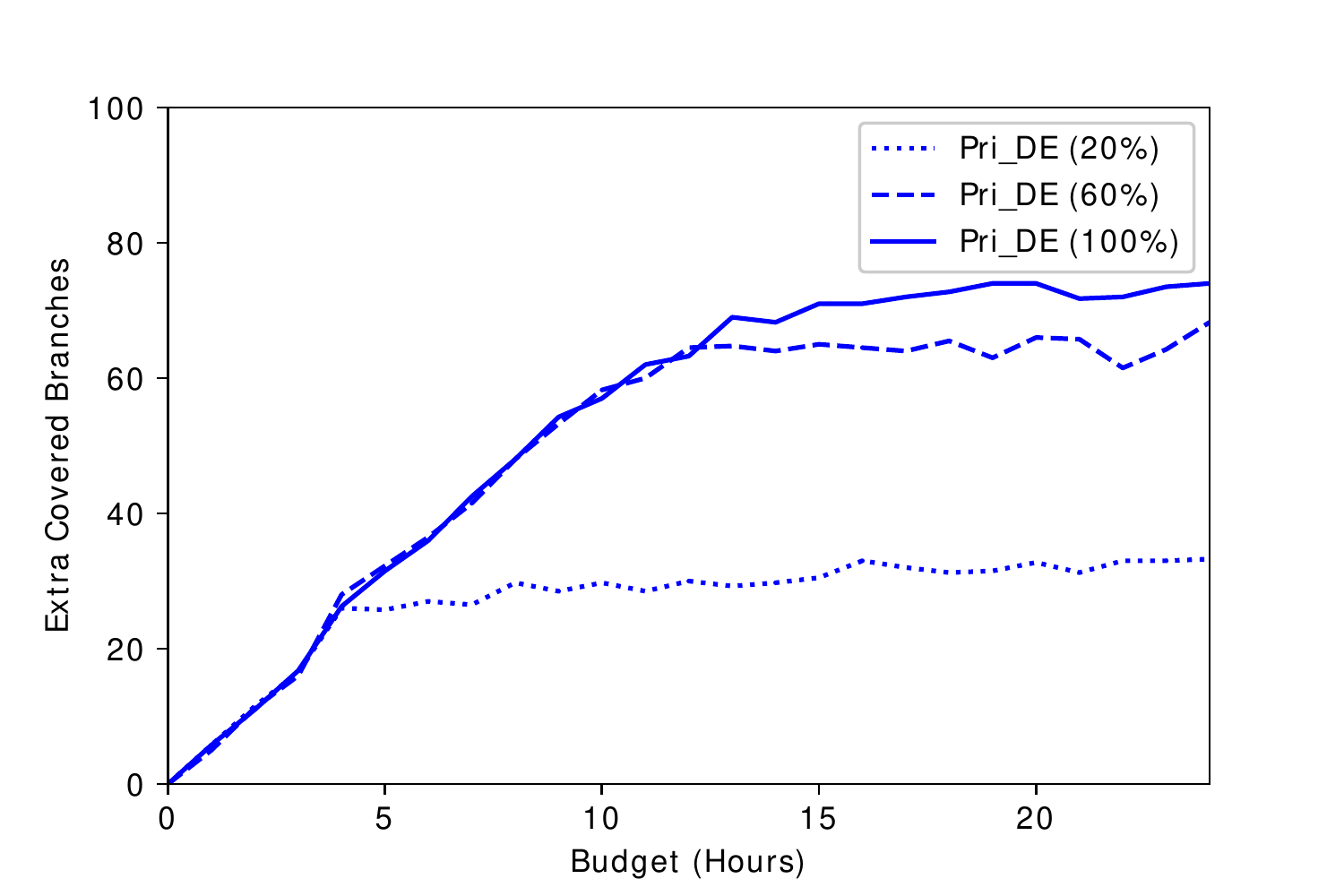}
\caption{Prioritized Subset Differential Evolution}
\label{fig:sub6}
\end{subfigure}
\caption{Extra Covered Branches vs. Tuning Budget for Different Subset Sizes, using six different class-level tuning. Cut-off values for the class selections are 20\%, 60\%, and 100\% (no selection). }
\label{fig:rq3}
\end{figure}

To answer this question, let's look at each of the six class-level tuning methods, one more time, when they are given more classes to tune, at Figure \ref{fig:rq3}. In this figure, each sub-graph plots one class-level method with three different cut-off values for the number of classes: 20\% (same as RQ2), 60\% the point where based on RQ1 from there onward the tuning gain is not increasing using our class prioritization heuristic, and 100\% which is the entire set of classes (in our test set).

The first  observation is that in all cases 60\% cut-off is much better than 20\% and in the prioritized versions (see sub-Figures \ref{fig:sub2}, \ref{fig:sub4}, and \ref{fig:sub6}) 60\% is almost as good as 100\%, with respect to the gained coverage. This matches with the findings of RQ1.1, where we saw our prioritization approach requires around 60\% of classes to reach the maximum gain. Note that the 20\% cut-off was based on previous studies and took into account a generic project's characteristics (where many classes are not tunable), whereas the 60\% cut-off is defined by an experiment on the training set of this study (where we excluded some easy classes in the beginning; thus a bigger portion of remaining classes are tunable). Therefore, the 60\% cut-off better reflects our test dataset. In practice, we suggest one runs a study similar to RQ1.1, once in the beginning, to find the best tuning cut-off for their project and always use that going forward.

We also notice that for non-prioritized versions (see sub-Figures \ref{fig:sub1}, \ref{fig:sub3}, and \ref{fig:sub5}), the coverage results from 60\% cut-off are lower than those of 100\%. This justifies the fact that prioritization helps find the best 60\% whereas a random subset of size 60\% might not be necessarily enough and more classes would be needed to tune, to cover all the tunable classes.

Additionally, in Meta-GA, we can observe that all the graphs have two consequent phases: an elevation phase and an almost plateau phase. In the elevation phase, Meta-GA is exploring more classes, in its first iteration. Covering more classes constantly increases the coverage. When the graph hits a plateau, all tunable classes have had at least one iteration of Meta-GA and now Meta-GA starts its exploitation (more iterations). The fewer the number of classes, the sooner the graph hits the plateau. Therefore, for Meta-GA if the budget is low, it is very probable that different selection sizes cover the same, as they both may be in the exploration phase. However, as we increase the budget, higher selection sizes cover more branches.

We also see exact same findings for DE, which matches observations of RQ2.2. that MG and DE's differences are insignificant. Thus in RQ4 and RQ5 we will not study DE as a baseline, anymore. 

Random Search's behavior, however, is different. Given any budget, search on 100 classes covers more branches than search on 20 classes. For example, given 10 hours of budget, random search on 100 classes covers almost twice as many as random search on 20 classes covers. Therefore, in this optimization problem, exploration is prior to exploitation.

\begin{tcolorbox}[drop shadow,enhanced,sharp corners,rounded corners=downhill,enlarge top initially by=5mm]
The optimum choice for subset size (i.e., prioritization cut-off) depends on the characteristics of classes under study and it needs to be determined using the static metrics proposed in RQ1.
\end{tcolorbox}

\subsubsection{RQ4 Results: Does reducing the search space, by selecting a subset of hyper-parameters, improve the effectiveness of tuning?}

An common observation from RQ2 and RQ3 is that our best approach (Pri\_MG) even on 100\% of classes cannot find the best configurations (e.g., see Figure \ref{fig:rq2}). The most common reason for not finding the optima might be to do with the search budget. Any evolutionary algorithm requires some iterations to find the optimum solution. However, our 24H budget is only enough to do two iterations. Therefore, in RQ4 and 5 we assign more budget to the Meta-GA per class. We do it in two ways. In RQ4, we use the same budget range (1H to 24H), but we reduce the search space. That makes finding the optimum easier for Meta-GA. In RQ5, however, we simply give more search time to meta-GA .

{\it RQ4.1: Which hyper-parameters are safer to discard, if we reduce the search space to a subset of hyper-parameters?}
As discussed in the design section, to reduce the search space we dropped the insignificant hyper-parameters, one at a time, to make three search spaces: large (including 4 hyper-parameters), medium (with 3 hyper-parameters), and small (with only 2 hyper-parameters). Running our feature selection on 100 random training sets (created by 100 random train-test splits), each containing 150 classes, we found the four, three, and two most important hyper-parameters, per class. Then, for each split, we found the hyper-parameters that were most frequently introduced as important hyper-parameters. 
Table \ref{tab:rq4.1} lists the frequency of each hyper-parameter being omitted, in 100 splits.

\begin{table}[h!]
    \centering
    \resizebox{\textwidth}{!}{\begin{tabular}{||c|ccccc||}
    \hline
        Search Space Size & Crossover Rate & Population Size & Elitism & Selection Function & Parent Check\\
        \hline
        \hline
        Large & 15 & 0 & 0 & 5 & \textbf{80} \\
        Medium & \textbf{60} & 5 & 10 & 25 & - \\
        Small & - & 20 & 25 & \textbf{55} & - \\
        \hline
    \end{tabular}}
    \caption{The frequency of each hyper-parameter being eliminated over 100 random splits, in Large, Medium, and Small search spaces.}
    \label{tab:rq4.1}
\end{table}

Given the results, the first hyper-parameter to exclude (least significant one) is ``Parent Check'', followed by ``Crossover Rate'', and then the ``Selection function''. Therefore, the most important hyper-parameters for the SBTG tool are the ``Population Size'' and ``Elitism'' (which is correlated to ``Population Size'' in some cases - for example, when the elitism rate is 10\% of the ``Population Size''). 

{\it RQ4.2: How does eliminating less important hyper-parameters affect the effectiveness of tuning?}
To evaluate the effect of reducing the search space, we use our best approach so far, Pri\_MG, on 60\% of classes, since 20\% cut-off was shown to be inadequate and selecting 60\% of classes was almost as good as 100\%, but leaving more budget for the search algorithm, per class.

\begin{figure}[h!]
\centering
\includegraphics[width=\linewidth]{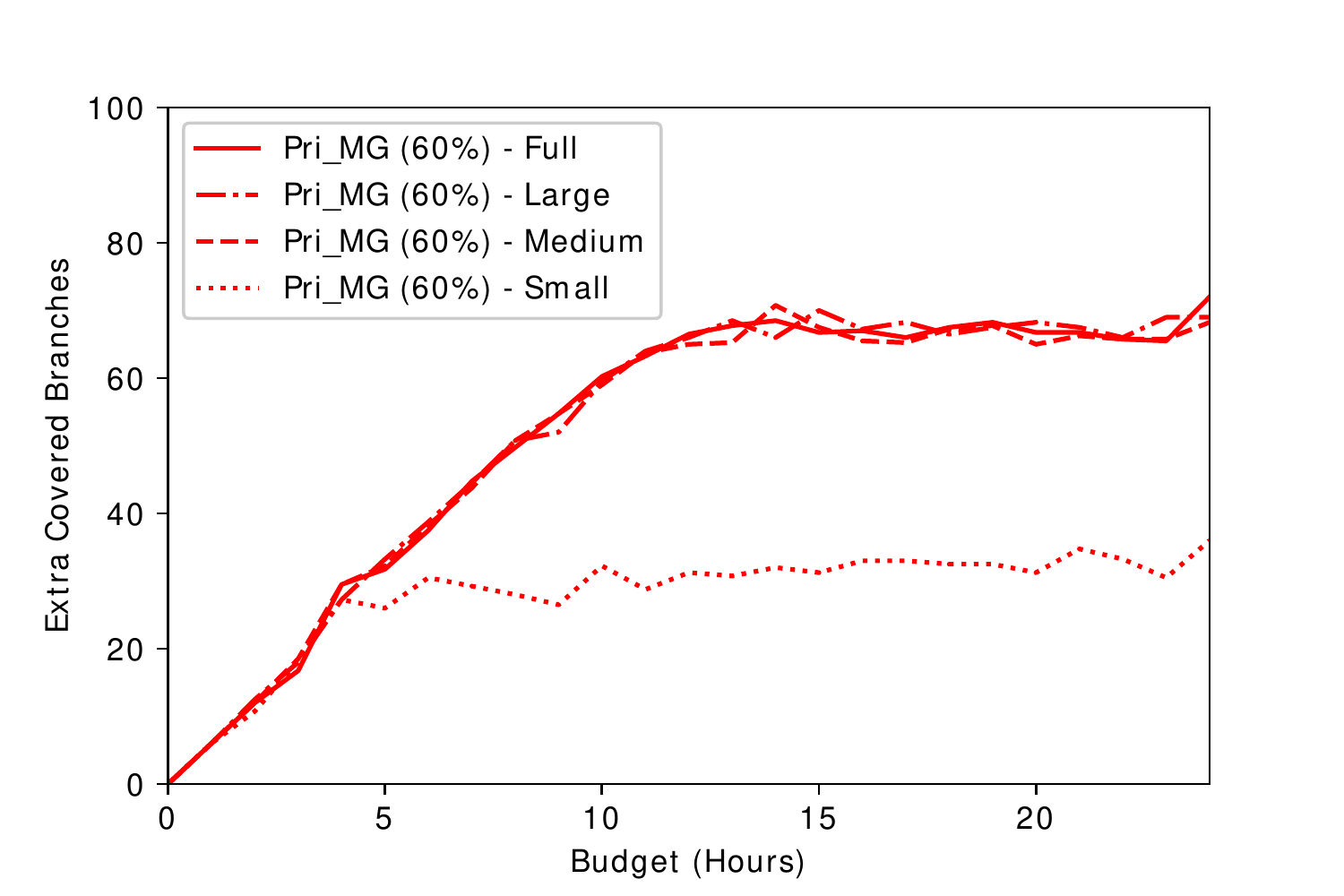}
\caption{Extra Covered Branches vs. Tuning Budget for search spaces with different sizes, tuned by Pri\_MG  using a 60\% cut-off.}
\label{fig:rq4}
\end{figure}

Figure \ref{fig:rq4}, shows the results of tuning with Pri\_MG (cut-off 60\%) on the initial search space as well the other three search spaces explained in RQ4.1. Note that for search spaces that use fewer hyper-parameters, the eliminated ones use their default values and are not being tuned anymore.  

Reducing the search space can have two effects on the search process: (1) since search space is getting smaller, there would be higher chance for a search method to find the optimal value within the smaller space, and (2) although, we try to eliminate only insignificant hyper-parameters from the search space, but we may still lose a few optimal points from the original space. The first effect potentially increases the final coverage while the second effect may reduce it. Therefore, in this experiment, we want to see which effect is stronger, or in other words, whether reducing the search space results in higher coverage or not.

It is observed from Figure \ref{fig:rq4} that eliminating the first two hyper-parameters (i.e. ``Parent Check'' and ``Crossover Rate'') that lead us to Large and Medium sizes did not change the search results, all with average extra covered branches of around 52. However, using the Small size search space, where ``Selection Function'' is also eliminated from the hyper-parameters, the coverage results are quite lower (average extra covered branches of 27.81) than in other search spaces.

Given that, we can conclude that in the Large and Medium search spaces most of optimal points are preserved in the search space, or we can find other similar optimal points when searching smaller spaces, but we don't find them sooner. Nonetheless, we are safe not to tune ``Parent Check'' and ``Crossover Rate'', anymore, in this experiment. Though this reduction, did not help improving our results but in other cases it might indeed lead to a faster convergence to an optimum solution. 

However, our results does not recommend dropping the ``Selection Function'', which could lead to much weaker results compared to the other search spaces. In addition, in the Small size search space, the only two hyper-parameters that are left are ``Population Size'' and ``Elitism Rate'' which also correlates to Population Size. Therefore, we conclude that tuning the ``Population Size'' for Search-based Test Generation Tools is a must and can significantly impact the results.

\begin{tcolorbox}[drop shadow,enhanced,sharp corners,rounded corners=downhill,enlarge top initially by=5mm]
Making the search space smaller does not lead to finding the best configurations sooner but it performs as good as searching over the entire search space. Therefore, eliminating the insignificant hyper-parameters is safe.
\end{tcolorbox}

\subsubsection{RQ5 Results: What is the effect of tuning budget on the effectiveness of our approach?}

\begin{figure}
\centering
\includegraphics[width=\linewidth]{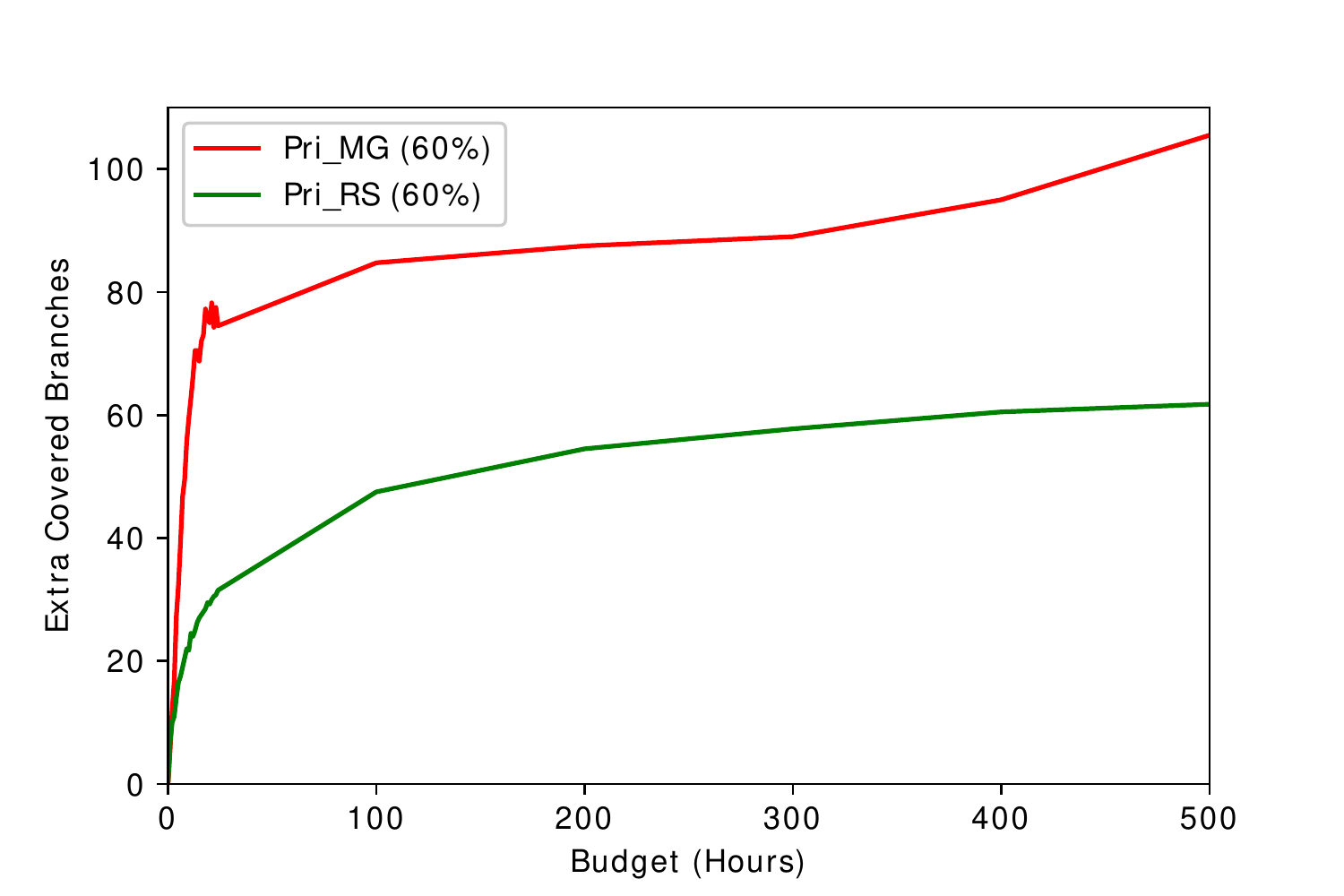}
\caption{Comparison of Pri\_MG and Pri\_RS, both using a 60\% cut-off, with time budgets more than 24 hours}
\label{fig:rq5}
\end{figure}

In this RQ we first see the effect of more tuning budget on the effectiveness of our approach while being applied on the original search space. Figure \ref{fig:rq5} compares Pri\_MG and Pri\_RS (as a baseline) results with up to 500H of tuning budget. The graph shows that both Random search and Meta-GA benefit from more time. At 24 hours, Meta-GA covers 74.5 extra branches. If we let it search the search space for 500 hours 105.5 branches are found. That is a about 50\% improvement in extra coverage.  Similarly, Pri\_RS covers 31.5 branches at 24H and 61.75 branches at 500H, which is 96\% improvement. 

It also seems that with very high budget the rate of improvement is higher for Meta-GA. For example, the improvement rate of Pri\_RS and Pri\_MG, from 400H budget to 500H are 1.25 and 10.5 branches per 100H, respectively. This motivates us to also look at the effect of more tuning budget together with reducing the search space. That will be our last effort to find the maximum potentials of a tuning method.

\begin{figure}
\centering
\includegraphics[width=\linewidth]{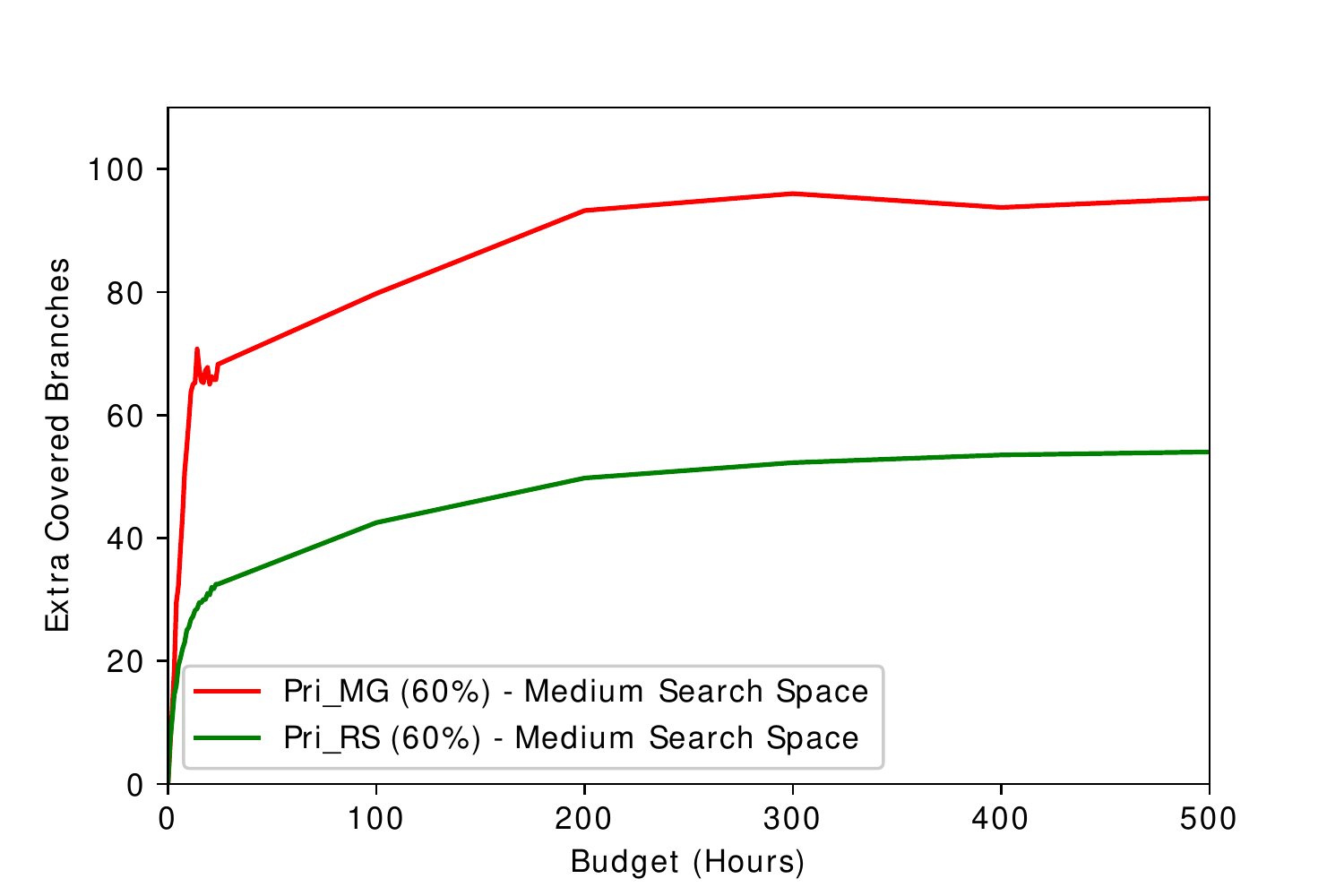}
\caption{Comparison of Pri\_MG and Pri\_RS, both using a 60\% cut-off, with time budgets more than 24 hours, in the Medium Search Space}
\label{fig:rq5_2}
\end{figure}

Figure \ref{fig:rq5_2}, however, shows that we can reach neither to the same coverage nor any faster with making the search space smaller.

In general, 500H for tuning is not practical. However, given that this task can be done, concurrently, for projects that definitely benefit from that extra 50\% coverage, it might worth to use extra computing in parallel and get the maximum benefit from tuning.

\begin{tcolorbox}[drop shadow,enhanced,sharp corners,rounded corners=downhill,enlarge top initially by=5mm]
The class prioritization component of our approach helps to tune for low budgets. To get the maximum from Meta-GA component of our approach though, a very high budget is needed. Whether the level of budget is justifiable is domain-dependant. In addition, the Meta-GA part can be parallelized that helps to reduce the tuning time.
\end{tcolorbox}

\subsection{Threats to Validity}
There are some potential threats to the construct validity of our results, as we used different metrics in our study and also our tuning approach is based on a newly defined measure. In order to show that our results correspond to what is supposed to be observed, we used measures like AUC and the number of covered branches that were previously used in the literature. In addition, besides the caution that we took to define a proper metric (Tuning Gain), we verified it by comparing its actual effect in terms of the well-defined branch coverage metric.

In order to mitigate conclusion threats, we repeated the methods that had a random nature, 25 times, we also repeated the prediction experiments with 100 train-test-splits and reported medians of the used metrics. When comparing different tuning methods in terms of their final code coverage, we also reported p\_Values of Mann–Whitney U test to make sure the differences are not due to the randomness nature of the algorithms. Finally, in order to show the effect of our tuning method in practice, we had a cost-benefit discussion in RQ3. 

To minimize internal validity, we stayed in-line with the prior research, in terms of using the same metrics and implementations that were available online. We used the static metrics that were previously used in other papers, and in order to calculate them, we used the open-source implementations that were used before in the literature. The subject under study was selected from the open-source benchmarks that are publicly available and were previously used in the literature. The artifact is also available online for the researchers to repeat our results. 

Regarding external validity, in order to show that our results of tuning can be generalized, we picked our classes out of 19 projects from three different sources, to reduce the bias to a project or benchmark dataset. However, all our experiments are run on EvoSuite, thus the results can not be generalized to other SBST tools, without further studies. In addition, the results for tuning are only available for Meta-GA, DE, and Random Search methods, and we cannot generalized these results to the results of other tuning methods without experiments.

\section{Future Work}
\label{future}

Several aspects of this thesis deserve more attention in the future, as follows:

(1) This work was initiated from replicating another paper in search-based testing. Therefore, as we wanted to compare our results with the baseline paper, we had to keep most of their experiments' design. All these tuning methods are applied to a grid of hyper-parameter, and the search space is discrete. For the future, we can change the design so that we can experiment more values per hyper-parameter and also add other hyper-parameters, as well. Some parameters seem to be important in Genetic Algorithm like Mutation Rate, but due to the above limitations, our experiments were not scalable to include other hyper-parameters.

(2) In this paper, we just focused on tuning the hyper-parameters of the EvoSuite tool. For the future, we can verify the results of this work by experimenting with other search-based test generation tools. Additionally, even for other test generation tools that do not use search-based methods, and use random search (such as Randoop), for example we can employ our prioritization technique to allocate the right amount of budget for each class, in test generation.

(3) The prioritization technique that we proposed is easy to apply to other search methods, however, in this paper, we just showed the result of prioritizing classes on three search methods, namely, Meta-GA, DE, and Random Search. Using other search methods may show the effectiveness of this approach better and may reduce the costs of tuning.

(4) In each optimization problem, plotting fitness values over the search space create a landscape, that is called the fitness landscape and the topology of this landscape gives us a better understanding of the optimization problem \cite{FLsurvey}. There are many properties defined for fitness landscapes. For example, if neighbouring solutions are very different from each other in terms of fitness value, the landscape is rugged, and otherwise is called smooth. Moreover, if the landscape has many flat areas, it is called neutral \cite{Rugged}. Knowing the degree of ruggedness, smoothness and neutrality of a landscape helps us find the fittest value easier. Here, we used Java classes' static features both to predict Tuning Gains and to shrink the search space. Another usage that one may follow in the future of this work is to predict the properties of the fitness landscape of each class. We can then cluster different types of fitness landscapes and define precise instructions for tuning each type of fitness landscape.

\section{Conclusion}
\label{conclusion}
Tuning evolutionary algorithms, in general, is known to be useful in many tasks. However, in the context of search-based test case generation (SBTG), tuning has not been much effective, so far. One of the reasons is that many classes in a project are not very tunable (they always get very high or very low coverage, no matter what configuration to use). Thus the tuning budget is wasted on these classes. In this paper, using static metrics to estimate each class's ``Tuning Gain'', we provided a class prioritization approach for tuning that allocates the tuning budget to the best subset of classes. We showed that our tuning approach can help the SBTG tool to generate cases that cover more branches compared to the test suite that was generated before tuning.


%
%


\printbibliography

@inproceedings{adaptive,
 author = {Xu, Xiong and Zhu, Ziming and Jiao, Li},
 title = {An Adaptive Fitness Function Based on Branch Hardness for Search Based Testing},
 booktitle = {Proceedings of the Genetic and Evolutionary Computation Conference},
 series = {GECCO '17},
 year = {2017},
%  isbn = {978-1-4503-4920-8},
 location = {Berlin, Germany},
 pages = {1335--1342},
 numpages = {8},
%  url = {http://doi.acm.org/10.1145/3071178.3071184},
%  doi = {10.1145/3071178.3071184},
 acmid = {3071184},
 publisher = {ACM},
 address = {New York, NY, USA},
}

@article{branchexpect,
title = "Estimating software testing complexity",
journal = "Information and Software Technology",
volume = "55",
number = "12",
pages = "2125 - 2139",
year = "2013",
% issn = "0950-5849",
% doi = "https://doi.org/10.1016/j.infsof.2013.07.007",
% url = "http://www.sciencedirect.com/science/article/pii/S0950584913001535",
author = "Javier Ferrer and Francisco Chicano and Enrique Alba",
}

@article{cov_predict,
author = {Grano, Giovanni and Titov, Timofey V. and Panichella, Sebastiano and Gall, Harald C.},
title = {Branch coverage prediction in automated testing},
journal = {Journal of Software: Evolution and Process},
volume = {31},
number = {9},
pages = {e2158},
keywords = {automated software testing, coverage prediction, machine learning, software testing},
% doi = {10.1002/smr.2158},
% url = {https://onlinelibrary.wiley.com/doi/abs/10.1002/smr.2158},
% eprint = {https://onlinelibrary.wiley.com/doi/pdf/10.1002/smr.2158},
note = {e2158 smr.2158},
abstract = {Abstract Software testing is crucial in continuous integration (CI). Ideally, at every commit, all the test cases should be executed, and moreover, new test cases should be generated for the new source code. This is especially true in a Continuous Test Generation (CTG) environment, where the automatic generation of test cases is integrated into the continuous integration pipeline. In this context, developers want to achieve a certain minimum level of coverage for every software build. However, executing all the test cases and, moreover, generating new ones for all the classes at every commit is not feasible. As a consequence, developers have to select which subset of classes has to be tested and/or targeted by test-case generation. We argue that knowing a priori the branch coverage that can be achieved with test-data generation tools can help developers into taking informed decision about those issues. In this paper, we investigate the possibility to use source-code metrics to predict the coverage achieved by test-data generation tools. We use four different categories of source-code features and assess the prediction on a large data set involving more than 3'000 Java classes. We compare different machine learning algorithms and conduct a fine-grained feature analysis aimed at investigating the factors that most impact the prediction accuracy. Moreover, we extend our investigation to four different search budgets. Our evaluation shows that the best model achieves an average 0.15 and 0.21 MAE on nested cross-validation over the different budgets, respectively, on EVOSUITE and RANDOOP. Finally, the discussion of the results demonstrate the relevance of coupling-related features for the prediction accuracy.},
year = {2019}
}

@InProceedings{khodam,
author="Zamani, Shayan
and Hemmati, Hadi",
editor="Nejati, Shiva
and Gay, Gregory",
title="Revisiting Hyper-Parameter Tuning for Search-Based Test Data Generation",
booktitle="Search-Based Software Engineering",
year="2019",
publisher="Springer International Publishing",
address="Cham",
pages="137--152",
% isbn="978-3-030-27455-9"
}

@Article{andrea,
author="Arcuri, Andrea
and Fraser, Gordon",
title="Parameter tuning or default values? An empirical investigation in search-based software engineering",
journal="Empirical Software Engineering",
year="2013",
month="June",
day="01",
volume="18",
number="3",
pages="594--623",
% issn="1573-7616",
% doi="10.1007/s10664-013-9249-9",
% url="https://doi.org/10.1007/s10664-013-9249-9"
}

@manual{cktool,
  title={Java code metrics calculator (CK)},
  author={Maurício Aniche},
  year={2015},
  note={Available at: \\ https://github.com/mauricioaniche/ck/}
}

@book{rsm,
 author = {Myers, Raymond H. and Montgomery, Douglas C.},
 title = {Response Surface Methodology: Process and Product in Optimization Using Designed Experiments},
 year = {1995},
%  isbn = {0471581003},
 edition = {1st},
 publisher = {John Wiley \& Sons, Inc.},
}

@book{frace,
 author = {Birattari, Mauro},
 title = {Tuning Metaheuristics: A Machine Learning Perspective},
 year = {2009},
%  isbn = {3642004822, 9783642004827},
 edition = {1st ed. 2005. 2nd printing},
 publisher = {Springer Publishing Company, Incorporated},
}

@article{controlGA,
 author = {Grefenstette, John},
 title = {Optimization of Control Parameters for Genetic Algorithms},
 journal={IEEE Transactions on Systems, Man, and Cybernetics},
 issue_date = {Jan./Feb. 1986},
 volume = {16},
 number = {1},
 month = jan,
 year = {1986},
%  issn = {0018-9472},
 pages = {122--128},
 numpages = {7},
%  url = {http://dx.doi.org/10.1109/TSMC.1986.289288},
%  doi = {10.1109/TSMC.1986.289288},
 acmid = {14123},
 publisher = {IEEE Press},
}

@INPROCEEDINGS{comparison, 
author={S. K. Smit and A. E. Eiben}, 
booktitle={2009 IEEE Congress on Evolutionary Computation}, 
title={Comparing parameter tuning methods for evolutionary algorithms}, 
year={2009}, 
volume={}, 
number={}, 
pages={399--406}, 
keywords={evolutionary computation;evolutionary algorithms;parameter tuning methods;human problem solving;Evolutionary computation;Iterative algorithms;Algorithm design and analysis;Genetic mutations;Humans;Fellows;Optimization methods;Traveling salesman problems;Cities and towns;Design methodology;evolutionary algorithms;parameter tuning}, 
% doi={10.1109/CEC.2009.4982974}, 
% ISSN={1089-778X}, 
month={May},}

@INPROCEEDINGS{toolcompetition, 
author={U. Rueda Molina and F. Kifetew and A. Panichella}, 
booktitle={2018 IEEE/ACM 11th International Workshop on Search-Based Software Testing (SBST)}, 
title={Java Unit Testing Tool Competition - Sixth Round}, 
year={2018}, 
volume={}, 
number={}, 
pages={22--29}, 
keywords={benchmark testing;Java;program testing;JUnit tool competitions;statistical analyses;automated JUnit testing tools;software projects;Java Unit Testing Tool Competition;performance assessment;benchmark projects;Tools;Benchmark testing;Java;Software;Measurement;Libraries;tool competition;benchmark;mutation testing;automation;unit testing;Java;statistical analysis;combined performance}, 
% doi={}, 
% ISSN={}, 
month={May},}

@InProceedings{GAbyGA,
author="Freisleben, Bernd
and H{\"a}rtfelder, Michael",
title="Optimization of Genetic Algorithms by Genetic Algorithms",
booktitle="Artificial Neural Nets and Genetic Algorithms",
year="1993",
publisher="Springer Vienna",
pages="392--399",
abstract="This paper presents an approach to determine the optimal Genetic Algorithm (GA), i.e. the most preferable type of genetic operators and their parameter settings, for a given problem. The basic idea is to consider the search for the best GA as an optimization problem and use another GA to solve it. As a consequence, a primary GA operates on a population of secondary GAs which in turn solve the problem in discussion. The paper describes how to encode the relevant information about GAs in gene strings and analyzes the impact of the individual genes on the results produced. The feasibility of the approach is demonstrated by presenting a parallel implementation on a multi-transputer system. Performance results for finding the best GA for the problem of optimal weight assignment in feed forward neural networks are presented.",
% isbn="978-3-7091-7533-0"
}

@article{sbst,
 author = {McMinn, Phil},
 title = {Search-based Software Test Data Generation: A Survey: Research Articles},
 journal = {Softw. Test. Verif. Reliab.},
 issue_date = {June 2004},
 volume = {14},
 number = {2},
 month = jun,
 year = {2004},
%  issn = {0960-0833},
 pages = {105--156},
 numpages = {52},
%  url = {http://dx.doi.org/10.1002/stvr.v14:2},
%  doi = {10.1002/stvr.v14:2},
 acmid = {1077279},
 publisher = {John Wiley and Sons Ltd.},
 keywords = {automated software test data generation, evolutionary algorithms, evolutionary testing, metaheuristic search, search-based software engineering, simulated annealing},
}

@article{sbse,
title = "Search-based software engineering",
journal = "Information and Software Technology",
volume = "43",
number = "14",
pages = {833--839},
year = "2001",
% issn = "0950-5849",
% doi = "https://doi.org/10.1016/S0950-5849(01)00189-6",
% url = "http://www.sciencedirect.com/science/article/pii/S0950584901001896",
author = "Mark Harman and Bryan F Jones",
keywords = "Software engineering, Metaheuristic, Genetic algorithm",
abstract = "This paper claims that a new field of software engineering research and practice is emerging: search-based software engineering. The paper argues that software engineering is ideal for the application of metaheuristic search techniques, such as genetic algorithms, simulated annealing and tabu search. Such search-based techniques could provide solutions to the difficult problems of balancing competing (and some times inconsistent) constraints and may suggest ways of finding acceptable solutions in situations where perfect solutions are either theoretically impossible or practically infeasible. In order to develop the field of search-based software engineering, a reformulation of classic software engineering problems as search problems is required. The paper briefly sets out key ingredients for successful reformulation and evaluation criteria for search-based software engineering."
}

@article{benchmark,
 author = {Fraser, Gordon and Arcuri, Andrea},
 title = {A Large-Scale Evaluation of Automated Unit Test Generation Using EvoSuite},
 journal = {ACM Transactions on Software Engineering and Methodology},
 issue_date = {December 2014},
 volume = {24},
 number = {2},
 month = dec,
 year = {2014},
%  issn = {1049-331X},
 pages = {8:1--8:42},
 articleno = {8},
 numpages = {42},
%  url = {http://doi.acm.org/10.1145/2685612},
%  doi = {10.1145/2685612},
 acmid = {2685612},
 publisher = {ACM},
 keywords = {JUnit, Java, Unit testing, automated test generation, benchmark, branch coverage, empirical software engineering},
}

@inproceedings{evosuite,
 author = {Fraser, Gordon and Arcuri, Andrea},
 title = {EvoSuite: Automatic Test Suite Generation for Object-oriented Software},
 booktitle = {Proceedings of the 19th ACM SIGSOFT Symposium and the 13th European Conference on Foundations of Software Engineering},
 year = {2011},
%  isbn = {978-1-4503-0443-6},
 numpages = {4},
%  url = {http://doi.acm.org/10.1145/2025113.2025179},
%  doi = {10.1145/2025113.2025179},
 acmid = {2025179},
 publisher = {ACM},
 keywords = {assertion generation, search based soft- ware testing, test case generation},
}

@inproceedings{harmancurrent,
  title={The current state and future of search based software engineering},
  author={Harman, Mark},
  booktitle={Future of Software Engineering (FOSE '07)},
  pages={342--357},
  year={2007},
}

@incollection{review1,
author={Edmund K. Burke and Mathew R. Hyde and Graham Kendall and Gabriela Ochoa and Ender Ozcan and John R. Woodward},
title = {Exploring Hyper-heuristic Methodologies with Genetic Programming},
booktitle={Computational Intelligence: Collaboration, Fusion and Emergence},
year="2009",
publisher="Springer Berlin Heidelberg",
pages="177--201",
abstract="Hyper-heuristics represent a novel search methodology that is motivated by the goal of automating the process of selecting or combining simpler heuristics in order to solve hard computational search problems. An extension of the original hyper-heuristic idea is to generate new heuristics which are not currently known. These approaches operate on a search space of heuristics rather than directly on a search space of solutions to the underlying problem which is the case with most meta-heuristics implementations. In the majority of hyper-heuristic studies so far, a framework is provided with a set of human designed heuristics, taken from the literature, and with good measures of performance in practice. A less well studied approach aims to generate new heuristics from a set of potential heuristic components. The purpose of this chapter is to discuss this class of hyper-heuristics, in which Genetic Programming is the most widely used methodology. A detailed discussion is presented including the steps needed to apply this technique, some representative case studies, a literature review of related work, and a discussion of relevant issues. Our aim is to convey the exciting potential of this innovative approach for automating the heuristic design process.",
% isbn="978-3-642-01799-5",
% doi="10.1007/978-3-642-01799-5_6",
% url="https://doi.org/10.1007/978-3-642-01799-5_6"
}

@article{review2,
title = "Parameter tuning of a choice-function based hyperheuristic using Particle Swarm Optimization",
journal = "Expert Systems with Applications",
volume = "40",
number = "5",
pages = "1690--1695",
year = "2013",
% issn = "0957-4174",
% doi = "https://doi.org/10.1016/j.eswa.2012.09.013",
% url = "http://www.sciencedirect.com/science/article/pii/S0957417412010676",
author = "Broderick Crawford and Ricardo Soto and Eric Monfroy and Wenceslao Palma and Carlos Castro and Fernando Paredes",
keywords = "Combinatorial optimization, Constraints satisfaction, Hyperheuristics, Particle Swarm",
abstract = "A Constraint Satisfaction Problem is defined by a set of variables and a set of constraints, each variable has a nonempty domain of possible values. Each constraint involves some subset of the variables and specifies the allowable combinations of values for that subset. A solution of the problem is defined by an assignment of values to some or all of the variables that does not violate any constraints. To solve an instance, a search tree is created and each node in the tree represents a variable of the instance. The order in which the variables are selected for instantiation changes the form of the search tree and affects the cost of finding a solution. In this paper we explore the use of a Choice Function to dynamically select from a set of variable ordering heuristics the one that best matches the current problem state in order to show an acceptable performance over a wide range of instances. The Choice Function is defined as a weighted sum of process indicators expressing the recent improvement produced by the heuristic recently used. The weights are determined by a Particle Swarm Optimization algorithm in a multilevel approach. We report results where our combination of strategies outperforms the use of individual strategies."
}

@inproceedings{replication,
 author = {Kotelyanskii, Anton and Kapfhammer, Gregory M.},
 title = {Parameter Tuning for Search-Based Test-Data Generation Revisited: Support for Previous Results},
 booktitle = {Proceedings of the 2014 14th International Conference on Quality Software},
 year = {2014},
%  isbn = {978-1-4799-7198-5},
 pages = {79--84},
 numpages = {6},
%  url = {https://doi.org/10.1109/QSIC.2014.43},
%  doi = {10.1109/QSIC.2014.43},
 acmid = {2707990},
 publisher = {IEEE Computer Society},
 keywords = {parameter tuning, test data generation, empirical studies},
}

@inproceedings{review3,
 author = {Craenen, Bart G. W. and Eiben, Agoston E.},
 title = {Stepwise Adaption of Weights with Refinement and Decay on Constraint Satisfaction Problems},
 booktitle = {Proceedings of the 3rd Annual Conference on Genetic and Evolutionary Computation},
 year = {2001},
%  isbn = {1-55860-774-9},
 pages = {291--298},
 numpages = {8},
%  url = {http://dl.acm.org/citation.cfm?id=2955239.2955290},
 acmid = {2955290},
 publisher = {Morgan Kaufmann Publishers Inc.},
}

@inproceedings{Feldt:2000,
 author = {Feldt, Robert and Nordin, Peter},
 title = {Using Factorial Experiments to Evaluate the Effect of Genetic Programming Parameters},
 booktitle = {Proceedings of the European Conference on Genetic Programming},
 year = {2000},
%  isbn = {3-540-67339-3},
 pages = {271--282},
 numpages = {12},
%  url = {http://dl.acm.org/citation.cfm?id=646808.703947},
 acmid = {703947},
 publisher = {Springer-Verlag},
}

@inproceedings{replicate,
 author = {Sayyad, Abdel Salam and Goseva-Popstojanova, Katerina and Menzies, Tim and Ammar, Hany},
 title = {On Parameter Tuning in Search Based Software Engineering: A Replicated Empirical Study},
 booktitle = {Proceedings of Workshop on Replication in Empirical Software Engineering},
 year = {2013},
%  isbn = {978-0-7695-5121-0},
%  url = {https://doi.org/10.1109/RESER.2013.6},
%  doi = {10.1109/RESER.2013.6},
 acmid = {2552635},
 publisher = {IEEE Computer Society},
 keywords = {Parameter Tuning, Search Based Software Engineering, Software Product Lines},
}

@inproceedings{MLtuning,
 author = {Song, Liyan and Minku, Leandro L. and Yao, Xin},
 title = {The Impact of Parameter Tuning on Software Effort Estimation Using Learning Machines},
 booktitle = {Proceedings of the 9th International Conference on Predictive Models in Software Engineering},
 year = {2013},
%  isbn = {978-1-4503-2016-0},
 pages = {9:1--9:10},
 articleno = {9},
 numpages = {10},
%  url = {http://doi.acm.org/10.1145/2499393.2499394},
%  doi = {10.1145/2499393.2499394},
 acmid = {2499394},
 publisher = {ACM},
 keywords = {ensembles, machine learning, online learning, sensitivity to parameters, software effort estimation},
}

@article{defect,
title = "Tuning for software analytics: Is it really necessary?",
journal = "Information and Software Technology",
volume = "76",
pages = {135--146},
year = "2016",
% issn = "0950-5849",
% doi = "https://doi.org/10.1016/j.infsof.2016.04.017",
% url = "http://www.sciencedirect.com/science/article/pii/S0950584916300738",
author = "Wei Fu and Tim Menzies and Xipeng Shen",
}

@inproceedings{cloneDetection,
 author = {Wang, Tiantian and Harman, Mark and Jia, Yue and Krinke, Jens},
 title = {Searching for Better Configurations: A Rigorous Approach to Clone Evaluation},
 booktitle = {Proceedings of the 2013 9th Joint Meeting on Foundations of Software Engineering},
 year = {2013},
%  isbn = {978-1-4503-2237-9},
 pages = {455--465},
 numpages = {11},
%  url = {http://doi.acm.org/10.1145/2491411.2491420},
%  doi = {10.1145/2491411.2491420},
 acmid = {2491420},
 publisher = {ACM},
 keywords = {Clone Detection, Genetic Algorithms, SBSE},
}

@article{mccabe,
 author = {McCabe, T. J.},
 title = {A Complexity Measure},
 journal = {IEEE Trans. Softw. Eng.},
 issue_date = {July 1976},
 volume = {2},
 number = {4},
 month = jul,
 year = {1976},
%  issn = {0098-5589},
 pages = {308--320},
 numpages = {13},
%  url = {https://doi.org/10.1109/TSE.1976.233837},
%  doi = {10.1109/TSE.1976.233837},
%  acmid = {1313586},
 publisher = {IEEE Press},
 address = {Piscataway, NJ, USA},
}

@article{test-suite,
 author = {Fraser, Gordon and Arcuri, Andrea},
 title = {Whole Test Suite Generation},
 journal={IEEE Transactions on Software Engineering}, 
 issue_date = {February 2013},
 volume = {39},
 number = {2},
 month = feb,
 year = {2013},
%  issn = {0098-5589},
 pages = {276--291},
 numpages = {16},
%  url = {http://dx.doi.org/10.1109/TSE.2012.14},
%  doi = {10.1109/TSE.2012.14},
%  acmid = {2478706},
 publisher = {IEEE Press},
 keywords = {Software,Genetic algorithms,Search problems,Arrays,Genetic programming,Software testing,collateral coverage,Search-based software engineering,length,branch coverage,genetic algorithm,infeasible goal},
}

@book{parallel,
 author = {Tange, Ole},
 title = {GNU Parallel 2018},
 year = {2018},
%  isbn = {1387509888, 9781387509881},
 publisher = {Lulu.com},
}

@InProceedings{mutation,
author="Paterson, David
and Turner, Jonathan
and White, Thomas
and Fraser, Gordon",
editor="Barros, M{\'a}rcio
and Labiche, Yvan",
title="Parameter Control in Search-Based Generation of Unit Test Suites",
booktitle="Search-Based Software Engineering",
year="2015",
publisher="Springer International Publishing",
address="Cham",
pages="141--156",
% isbn="978-3-319-22183-0"
}

@article{panichella2017automated,
title={Automated test case generation as a many-objective optimisation problem with dynamic selection of the targets},
author={Panichella, Annibale and Kifetew, Fitsum Meshesha and Tonella, Paolo},
journal={IEEE Transactions on Software Engineering},
volume={44},
number={2},
pages={122--158},
year={2017},
publisher={IEEE}
}

@article{scikit-learn,
 title={Scikit-learn: Machine Learning in {P}ython},
 author={Pedregosa, F. and Varoquaux, G. and Gramfort, A. and Michel, V.
         and Thirion, B. and Grisel, O. and Blondel, M. and Prettenhofer, P.
         and Weiss, R. and Dubourg, V. and Vanderplas, J. and Passos, A. and
         Cournapeau, D. and Brucher, M. and Perrot, M. and Duchesnay, E.},
 journal={Journal of Machine Learning Research},
 volume={12},
 pages={2825--2830},
 year={2011}
}

@Inbook{Feurer2019,
author="Feurer, Matthias
and Hutter, Frank",
editor="Hutter, Frank
and Kotthoff, Lars
and Vanschoren, Joaquin",
title="Hyperparameter Optimization",
bookTitle="Automated Machine Learning: Methods, Systems, Challenges",
year="2019",
publisher="Springer International Publishing",
address="Cham",
pages="3--33",
isbn="978-3-030-05318-5",
doi="10.1007/978-3-030-05318-5_1",
url="https://doi.org/10.1007/978-3-030-05318-5_1"
}

@incollection{bookchapter_testing,
  author = {Gordon Fraser and Jos\'e Miguel Rojas},
  title = {Software Testing},
  editor = {Sungdeok Cha and Richard N. Taylor and Kyo C. Kang},
  booktitle = {Handbook of Software Engineering},
  publisher = {Springer International Publishing},
  year = 2018
}

@Inbook{RFE,
author="Guyon, Isabelle
and Elisseeff, Andr{\'e}",
editor="Guyon, Isabelle
and Nikravesh, Masoud
and Gunn, Steve
and Zadeh, Lotfi A.",
title="An Introduction to Feature Extraction",
bookTitle="Feature Extraction: Foundations and Applications",
year="2006",
publisher="Springer Berlin Heidelberg",
address="Berlin, Heidelberg",
pages="1--25"
}

@article{NCG, author = {J\"{a}rvelin, Kalervo and Kek\"{a}l\"{a}inen, Jaana}, title = {Cumulated Gain-Based Evaluation of IR Techniques}, year = {2002}, issue_date = {October 2002}, publisher = {Association for Computing Machinery}, address = {New York, NY, USA}, volume = {20}, number = {4}, issn = {1046-8188}, url = {https://doi.org/10.1145/582415.582418}, doi = {10.1145/582415.582418}, journal = {ACM Trans. Inf. Syst.}, month = oct, pages = {422–446}, numpages = {25} }

@misc{jdepend, title={Jdepend}, url={https://github.com/clarkware/jdepend}, journal={GitHub}, author={Clarkware}, year={2020}, month={Mar}}

@misc{silly, title={KeyWord Counter}, url={https://github.com/lukaseder/silly-metrics}, journal={GitHub}, author={Lukas Eder}}

@misc{halstead, title={Halstead ComplexityMeasures}, url={https://github.com/aametwally/Halstead-Complexity-Measures}, journal={GitHub}, author={Ahmed A. Metwally}}

@article{FLsurvey,
title = "A survey of techniques for characterising fitness landscapes and some possible ways forward",
journal = "Information Sciences",
volume = "241",
pages = "148 - 163",
year = "2013",
issn = "0020-0255",
doi = "https://doi.org/10.1016/j.ins.2013.04.015",
author = "Katherine M. Malan and Andries P. Engelbrecht",
}

@Inbook{Rugged,
author="Vassilev, Vesselin K.
and Fogarty, Terence C.
and Miller, Julian F.",
editor="Ghosh, Ashish
and Tsutsui, Shigeyoshi",
title="Smoothness, Ruggedness and Neutrality of Fitness Landscapes: from Theory to Application",
bookTitle="Advances in Evolutionary Computing: Theory and Applications",
year="2003",
publisher="Springer Berlin Heidelberg",
address="Berlin, Heidelberg",
pages="3--44",
isbn="978-3-642-18965-4",
doi="10.1007/978-3-642-18965-4_1",
url="https://doi.org/10.1007/978-3-642-18965-4_1"
}

@inproceedings{SMOTUNED,
author = {Agrawal, Amritanshu and Menzies, Tim},
title = {Is “Better Data” Better than “Better Data Miners”? On the Benefits of Tuning SMOTE for Defect Prediction},
year = {2018},
isbn = {9781450356381},
publisher = {Association for Computing Machinery},
address = {New York, NY, USA},
url = {https://doi.org/10.1145/3180155.3180197},
doi = {10.1145/3180155.3180197},
booktitle = {Proceedings of the 40th International Conference on Software Engineering},
pages = {1050–1061},
numpages = {12},
location = {Gothenburg, Sweden},
series = {ICSE ’18}
}

@article{differential_evolution,
author = {Storn, Rainer and Price, Kenneth},
title = {Differential Evolution – A Simple and Efficient Heuristic for Global Optimization over Continuous Spaces},
year = {1997},
issue_date = {December 1997},
publisher = {Kluwer Academic Publishers},
address = {USA},
volume = {11},
number = {4},
issn = {0925-5001},
url = {https://doi.org/10.1023/A:1008202821328},
doi = {10.1023/A:1008202821328},
journal = {J. of Global Optimization},
month = dec,
pages = {341–359},
numpages = {19}
}

@article{DE+RF,
  title={Why is differential evolution better than grid search for tuning defect predictors?},
  author={Fu, Wei and Nair, Vivek and Menzies, Tim},
  journal={arXiv preprint arXiv:1609.02613},
  year={2016}
}

@article{necessary,
author = {Fu, Wei and Menzies, Tim and Shen, Xipeng},
title = {Tuning for Software Analytics},
year = {2016},
issue_date = {August 2016},
publisher = {Butterworth-Heinemann},
address = {USA},
volume = {76},
number = {C},
issn = {0950-5849},
url = {https://doi.org/10.1016/j.infsof.2016.04.017},
doi = {10.1016/j.infsof.2016.04.017},
journal = {Inf. Softw. Technol.},
month = aug,
pages = {135–146},
numpages = {12}
}

@inproceedings{Fast-and-Frugal,
author = {Chen, Di and Fu, Wei and Krishna, Rahul and Menzies, Tim},
title = {Applications of Psychological Science for Actionable Analytics},
year = {2018},
isbn = {9781450355735},
publisher = {Association for Computing Machinery},
address = {New York, NY, USA},
url = {https://doi.org/10.1145/3236024.3236050},
doi = {10.1145/3236024.3236050},
booktitle = {Proceedings of the 2018 26th ACM Joint Meeting on European Software Engineering Conference and Symposium on the Foundations of Software Engineering},
pages = {456–467},
numpages = {12},
location = {Lake Buena Vista, FL, USA},
series = {ESEC/FSE 2018}
}

@article{topicModelling,
  author    = {Amritanshu Agrawal and
               Wei Fu and
               Tim Menzies},
  title     = {What is Wrong with Topic Modeling? (and How to Fix it Using Search-based
               {SE)}},
  journal   = {CoRR},
  volume    = {abs/1608.08176},
  year      = {2016},
  url       = {http://arxiv.org/abs/1608.08176},
  archivePrefix = {arXiv},
  eprint    = {1608.08176},
  bibsource = {dblp computer science bibliography, https://dblp.org}
}

@article{effortEstim,
  author    = {Tianpei Xia and
               Rahul Krishna and
               Jianfeng Chen and
               George Mathew and
               Xipeng Shen and
               Tim Menzies},
  title     = {Hyperparameter Optimization for Effort Estimation},
  journal   = {CoRR},
  volume    = {abs/1805.00336},
  year      = {2018},
  url       = {http://arxiv.org/abs/1805.00336},
  archivePrefix = {arXiv},
  eprint    = {1805.00336},
  bibsource = {dblp computer science bibliography, https://dblp.org}
}

@misc{improvedSecBug,
    title={Improved Recognition of Security Bugs via Dual Hyperparameter Optimization},
    author={Rui Shu and Tianpei Xia and Jianfeng Chen and Laurie Williams and Tim Menzies},
    year={2019},
    eprint={1911.02476},
    archivePrefix={arXiv},
    primaryClass={cs.SE}
}

@article{sbst_survey,
author = {Ali, Shaukat and Briand, Lionel C. and Hemmati, Hadi and Panesar-Walawege, Rajwinder Kaur},
title = {A Systematic Review of the Application and Empirical Investigation of Search-Based Test Case Generation},
year = {2010},
issue_date = {November 2010},
publisher = {IEEE Press},
volume = {36},
number = {6},
issn = {0098-5589},
url = {https://doi.org/10.1109/TSE.2009.52},
doi = {10.1109/TSE.2009.52},
journal = {IEEE Trans. Softw. Eng.},
month = nov,
pages = {742–762},
numpages = {21}
}

@inproceedings{DOTgEAr,
author = {Oster, Norbert and Saglietti, Francesca},
title = {Automatic Test Data Generation by Multi-Objective Optimisation},
year = {2006},
isbn = {3540457623},
publisher = {Springer-Verlag},
address = {Berlin, Heidelberg},
url = {https://doi.org/10.1007/11875567_32},
doi = {10.1007/11875567_32},
booktitle = {Proceedings of the 25th International Conference on Computer Safety, Reliability, and Security},
pages = {426–438},
numpages = {13},
location = {Gdansk, Poland},
series = {SAFECOMP’06}
}

@article{JTExpert,
  title={Instance generator and problem representation to improve object oriented code coverage},
  author={Sakti, Abdelilah and Pesant, Gilles and Gu{\'e}h{\'e}neuc, Yann-Ga{\"e}l},
  journal={IEEE Transactions on Software Engineering},
  volume={41},
  number={3},
  pages={294--313},
  year={2014},
  publisher={IEEE}
}

@inproceedings{algorithms_for_HPO,
author = {Bergstra, James and Bardenet, R\'{e}mi and Bengio, Yoshua and K\'{e}gl, Bal\'{a}zs},
title = {Algorithms for Hyper Parameter Optimization},
year = {2011},
isbn = {9781618395993},
publisher = {Curran Associates Inc.},
address = {Red Hook, NY, USA},
booktitle = {Proceedings of the 24th International Conference on Neural Information Processing Systems},
pages = {2546–2554},
numpages = {9},
location = {Granada, Spain},
series = {NIPS’11}
}

@inproceedings{classification_tuning,
  title={Automated parameter optimization of classification techniques for defect prediction models},
  author={Tantithamthavorn, Chakkrit and McIntosh, Shane and Hassan, Ahmed E and Matsumoto, Kenichi},
  booktitle={Proceedings of the 38th International Conference on Software Engineering},
  pages={321--332},
  year={2016}
}

@article{Eiben2003,
  title={Parameter control in evolutionary algorithms},
  author={Eiben, {\'A}goston E and Hinterding, Robert and Michalewicz, Zbigniew},
  journal={IEEE Transactions on evolutionary computation},
  volume={3},
  number={2},
  pages={124--141},
  year={1999},
  publisher={IEEE}
}

@inproceedings{thomas,
  title={Automatic Configuration of Deep Neural Networks with Parallel Efficient Global Optimization},
  author={van Stein, Bas and Wang, Hao and B{\"a}ck, Thomas},
  booktitle={2019 International Joint Conference on Neural Networks (IJCNN)},
  pages={1--7},
  year={2019},
  organization={IEEE}
}

@article{random_bengio,
author = {Bergstra, James and Bengio, Yoshua},
title = {Random Search for Hyper Parameter Optimization},
year = {2012},
issue_date = {3/1/2012},
publisher = {JMLR.org},
volume = {13},
number = {null},
issn = {1532-4435},
journal = {J. Mach. Learn. Res.},
month = feb,
pages = {281–305},
numpages = {25}
}

@book{goldberg2006genetic,
  title={Genetic algorithms},
  author={Goldberg, David E},
  year={2006},
  publisher={Pearson Education India}
}

@inproceedings{boa,
  title={BOA: The Bayesian optimization algorithm},
  author={Pelikan, Martin and Goldberg, David E and Cant{\'u}-Paz, Erick and others},
  booktitle={Proceedings of the genetic and evolutionary computation conference GECCO-99},
  volume={1},
  pages={525--532},
  year={1999},
  organization={Citeseer}
}

@article{simulated_annealing,
  title={Optimization by simulated annealing},
  author={Kirkpatrick, Scott and Gelatt, C Daniel and Vecchi, Mario P},
  journal={science},
  volume={220},
  number={4598},
  pages={671--680},
  year={1983},
  publisher={American association for the advancement of science}
}

@inproceedings{topicModel_GA,
  title={How to effectively use topic models for software engineering tasks? an approach based on genetic algorithms},
  author={Panichella, Annibale and Dit, Bogdan and Oliveto, Rocco and Di Penta, Massimilano and Poshynanyk, Denys and De Lucia, Andrea},
  booktitle={2013 35th International Conference on Software Engineering (ICSE)},
  pages={522--531},
  year={2013},
  organization={IEEE}
}

@article{dodge,
  author    = {Amritanshu Agrawal and
               Wei Fu and
               Di Chen and
               Xipeng Shen and
               Tim Menzies},
  title     = {How to "DODGE" Complex Software Analytics?},
  journal   = {CoRR},
  volume    = {abs/1902.01838},
  year      = {2019},
  url       = {http://arxiv.org/abs/1902.01838},
  archivePrefix = {arXiv},
  eprint    = {1902.01838},
  bibsource = {dblp computer science bibliography, https://dblp.org}
}

@article{ruhe_release_planning,
author = {Zhang, Yuanyuan and Harman, Mark and Ochoa, Gabriela and Ruhe, Guenther and Brinkkemper, Sjaak},
title = {An Empirical Study of Meta- and Hyper-Heuristic Search for Multi-Objective Release Planning},
year = {2018},
issue_date = {June 2018},
publisher = {Association for Computing Machinery},
address = {New York, NY, USA},
volume = {27},
number = {1},
issn = {1049-331X},
url = {https://doi.org/10.1145/3196831},
doi = {10.1145/3196831},
journal = {ACM Trans. Softw. Eng. Methodol.},
month = jun,
articleno = {3},
numpages = {32}
}

@book{differential_tuning,
  title={Differential evolution: a practical approach to global optimization},
  author={Price, Kenneth and Storn, Rainer M and Lampinen, Jouni A},
  year={2006},
  publisher={Springer Science \& Business Media}
}

\end{document}